\documentclass[12pt,letterpaper]{article}
\pdfoutput=1
\usepackage[colorlinks=true,
linkcolor=red, 
citecolor=red,
filecolor=red,
urlcolor=red,
linktoc=page, 
pdfstartview=FitV,
bookmarksopen=true]{hyperref}
\usepackage[left=2cm,top=1cm,right=3cm,nohead]{geometry}
\usepackage[dvipsnames]{xcolor}
\usepackage{colortbl,cancel}
\usepackage[utf8]{inputenc} 
\usepackage{color}
\usepackage{epsfig,latexsym,amsfonts,mathtools,amsthm,amssymb,amsbsy,multirow,slashed,
	mathrsfs,wasysym,textcomp,subfigure,wrapfig,comment,bbold,array,longtable,multirow,setspace,amsmath,pifont,nicefrac}
\usepackage{tabularx}
\usepackage[]{mdframed}
\usepackage{graphicx}
\usepackage{setspace}
\usepackage{subfigure}
\usepackage{multirow}
\usepackage[bf]{caption}
\usepackage{braket}
\usepackage{comment}
\usepackage{float}
\usepackage{tikz}
\usetikzlibrary{mindmap,trees,shadows}
\colorlet{color1}{gray!25}
\usetikzlibrary{positioning}
\usetikzlibrary{intersections} 
\newlength{\PicScale}
\usepackage[UKenglish]{babel}
\usepackage[toc,page]{appendix}
\usepackage{hhline}
\usepackage{geometry}
\usepackage{cite}
\usepackage{datetime}
\usepackage{bbm} 
\usepackage{bm} 
\hypersetup{
	pdftitle={},
	pdfauthor={},
	pdfsubject={}
}

\newcolumntype{M}[1]{>{\centering\arraybackslash}m{#1}}
\newcolumntype{N}{@{}m{0pt}@{}}

\numberwithin{equation}{section}

\makeatletter
\def\@cline#1-#2\@nil{
	\omit
	\@multicnt#1
	\advance\@multispan\m@ne
	\ifnum\@multicnt=\@ne\@firstofone{&\omit}\fi
	\@multicnt#2
	\advance\@multicnt-#1
	\advance\@multispan\@ne
	\leaders\hrule\@height\arrayrulewidth\hfill
	\cr
	\noalign{\nobreak\vskip-\arrayrulewidth}}
\makeatother

\begin{document}
\pagestyle{empty}
\begin{center}        
  {\bf\LARGE Non-supersymmetric heterotic strings\\ and chiral CFTs\\ [3mm]}

\large{ H\'ector Parra De Freitas
 \\[2mm]}

{\small  Jefferson Physical Laboratory, Harvard University\\ [-1mm]}
{\small\textit{Cambridge, MA 02138, USA}\\[0.2cm]}

{\small \verb"hparradefreitas@fas.harvard.edu"\\[-3mm]}
\vspace{0.3in}

\small{\bf Abstract} \\[3mm]\end{center}
Non-supersymmetric heterotic strings share various properties with their supersymmetric counterparts. Torus compactifications of the latter live in a component of the moduli space of string vacua with 16 supercharges, and various asymmetric orbifolds thereof realize vacua in other components, exhibiting qualitative differences such as rank reduction. We set out to study the analogous problem for non-supersymmetric heterotic strings, framing it in relation to chiral fermionic CFTs with central charge 24, which were classified recently. We find that for the case analogous to the so-called CHL string, which has gauge group rank reduced by 8, there are in total four non-supersymmetric versions. These include the well known $E_8$ string and three other constructions a la CHL, which can be distinguished qualitatively by how tachyons appear in their classical moduli spaces. We also discuss the classification problem for lower rank theories and the relationship between MSDS models and Scherk-Schwarz reductions.

\newpage



\setcounter{page}{1}
\pagestyle{plain}
\renewcommand{\thefootnote}{\arabic{footnote}}
\setcounter{footnote}{0}

\tableofcontents	
\newpage
\section{Introduction}
\label{s:intro}
String compactifications with broken supersymmetry are out of analytic control insofar as there are no BPS states on the one hand and on the other various quantities fail to cancel due to a generic lack of Bose-Fermi degeneracy in the spectrum. However, one can start from a supersymmetric setting and break supersymmetry in such a way that various properties of the original theory are preserved. One may then learn something about the novel phenomena tied to supersymmetry breaking by leveraging the control one had in the parent supersymmetric theory. 

\subsection{Non-supersymmetric heterotic strings}

A characteristic example of this scenario is provided by the non-supersymmetric heterotic strings\cite{Dixon:1986iz,Alvarez-Gaume:1986ghj,Kawai:1986vd,Seiberg:1986by}. Upon circle compactification, the supersymmetric $SO(32)$ and $E_8 \times E_8$ heterotic theories become dual \cite{Ginsparg:1986bx} and give rise to a Narain moduli space through the vevs of the compactification radius and Wilson lines \cite{Narain:1985jj,Narain:1986am}. Generic points in this space have Abelian gauge group $U(1)^{17}_L \times U(1)_R$ due to the Wilson lines, but at special loci there appear massless gauge bosons and the left moving contribution is enhanced up to a fully non-Abelian rank 17 gauge group. This full enhancement is a stringy phenomenon made possible by the presence of massless winding states. If one then breaks supersymmetry by imposing anti-periodic boundary conditions for the spacetime fermions along the compactification circle, it turns out that the aforementioned properties are preserved, although the detailed aspects get enriched.

\begin{figure}
	\centering
	\begin{tikzpicture}[scale = 1.25]
		\draw(0,0)--(7,0);
		\draw(0.5,0)--(0.5,1);	
		\draw(6.5,0)--(6.5,1);
		\draw[fill=white](0,0)circle(0.1);
		\draw[fill=white](0.5,0)circle(0.1);
		\draw[fill=white](1,0)circle(0.1);
		\draw[fill=white](1.5,0)circle(0.1);
		\draw[fill=white](2,0)circle(0.1);
		\draw[fill=white](2.5,0)circle(0.1);
		\draw[fill=white](3,0)circle(0.1);
		\draw[fill=white](3.5,0)circle(0.1);
		\draw[fill=white](4,0)circle(0.1);
		\draw[fill=white](4.5,0)circle(0.1);
		\draw[fill=white](5,0)circle(0.1);
		\draw[fill=white](5.5,0)circle(0.1);
		\draw[fill=white](6,0)circle(0.1);
		\draw[fill=white](6.5,0)circle(0.1);
		\draw[fill=white](7,0)circle(0.1);
		\draw[fill=white](0.5,0.5)circle(0.1);
		\draw[fill=white](0.5,1)circle(0.1);
		\draw[fill=white](6.5,0.5)circle(0.1);
		\draw[fill=white](6.5,1)circle(0.1);
		\begin{scope}[shift={(9,0)}]
			\draw(0,0)--(0.5,0.5);
			\draw(2,0)--(1.5,0.5);
			\draw(0,2)--(0.5,1.5);
			\draw(2,2)--(1.5,1.5);
			
			\draw[double, Red](0.5,0.5)--(1,0.75);
			\draw[double, Red](0.5,1.5)--(1,1.25);
			\draw[double, Red](1,0.75)--(1.5,1.5);
			\draw[double, Red, fill=white](1,1.25)--(1.5,0.5);
			\draw[double, Red](1,0.75)--(1,1.25);
			
			\draw(0,0)--(2,0)--(2,2)--(0,2)--(0,0);
			\draw[fill=white](0,0)circle(0.1);
			\draw[fill=white](0.5,0)circle(0.1);
			\draw[fill=white](1,0)circle(0.1);
			\draw[fill=white](1.5,0)circle(0.1);
			\draw[fill=white](2,0)circle(0.1);
			\draw[fill=white](0,0.5)circle(0.1);
			\draw[fill=white](0,1)circle(0.1);
			\draw[fill=white](0,1.5)circle(0.1);
			\draw[fill=white](2,0.5)circle(0.1);
			\draw[fill=white](2,1)circle(0.1);
			\draw[fill=white](2,1.5)circle(0.1);	
			\draw[fill=white](0,2)circle(0.1);
			\draw[fill=white](0.5,2)circle(0.1);
			\draw[fill=white](1,2)circle(0.1);
			\draw[fill=white](1.5,2)circle(0.1);
			\draw[fill=white](2,2)circle(0.1);
			
			\draw[fill=white](0.5,0.5)circle(0.1);
			\draw[fill=white](0.5,1.5)circle(0.1);
			\draw[fill=white](1.5,0.5)circle(0.1);
			\draw[fill=white](1.5,1.5)circle(0.1);
			\draw[Red, fill=white,double](1,1.25)circle(0.1);
			\draw[Red, fill=white,double](1,0.75)circle(0.1);
		\end{scope}
		
	\end{tikzpicture}
	\caption{Coxeter-Dynkin diagrams for the supersymmetric and non-supersymmetric heterotic strings of rank 16 compactified on the circle. Simple nodes correspond to reflective vetors with norm $2$ and (red) double nodes to reflective vectors with norm $1$. Simple links denote inner product $-1$ while (red) double links denote inner product $-2$. These root systems generate the group of reflections inside the full T-duality groups of the theories and encode every possible non-Abelian symmetry enhancement and infinite distance limit in the form of subdiagrams of respectively finite and affine type. Both diagrams were first discovered by Vinberg \cite{Vinberg} in the context of the classification of hyperbolic Coxeter groups. Their applications to heterotic strings were developed respectively in \cite{Goddard:1986bp,Cachazo:2000ey} and \cite{Fraiman:2023cpa}.}
	\label{fig:GDD}
\end{figure}
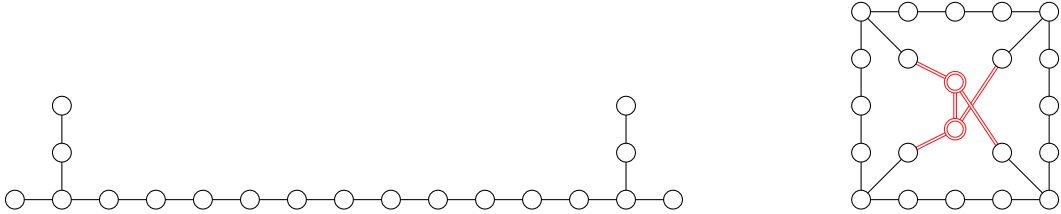

Breaking supersymmetry in this manner corresponds to an application of the Scherk-Schwarz mechanism \cite{Scherk:1979zr,Ferrara:1987es} in stringy form \cite{Kounnas:1989dk}. For supersymmetric heterotic strings, the form of the classical moduli space is preserved locally, i.e.
\begin{equation}
	\mathcal{M}_{\cancel{\text{SUSY}}}^\text{local} = \mathcal{M}_\text{SUSY}^\text{local} = O(17,1,\mathbb{R})/O(17,\mathbb{R})\,,
\end{equation} 
up to the dilaton contribution $\mathbb{R}$. The group of T-dualities, which defines the global form of the moduli space, is however modified. In both cases, the T-duality group is the automorphism group of the electric charge lattice of the theory. In the supersymmetric case this is the usual even self-dual Narain lattice $\Gamma_N$ while in the non-supersymmetric case it is the dual of an index 2 sublattice of $\Gamma_N$. The resulting T-duality groups are of hyperbolic Coxeter form, and were obtained a long time ago by Vinberg in \cite{Vinberg}. The reflections in these groups are captured by Coxeter-Dynkin diagrams, depicted in Figure \ref{fig:GDD}, which yield every possible non-Abelian symmetry enhancement as well as every infinite distance limit. In the first case these limits are decompactifications to the two ten dimensional supersymmetric heterotic strings while in the second case there are eight limits yielding all of the eight heterotic strings with rank 16 \cite{Ginsparg:1986wr,Itoyama:1986ei}. Indeed, the Scherk-Schwarz mechanism applied to the supersymmetric heterotic strings on a circle is T-dual to the circle compactification of the non-supersymmetric heterotic strings of rank 16, e.g. the $O(16)\times O(16)$ string, and the discussion presented so far indicates a resemblance between the two types of heterotic strings.

As supersymmetry is broken, the resulting theory no longer enjoys Bose-Fermi degeneracy hence the vacuum energy $\Lambda$ is generically nonzero, and certain regions in the classical moduli space become populated by tachyons. As was shown in \cite{Ginsparg:1986wr}, maximal enhancements in the classical moduli space extremize $\Lambda$ to all orders in perturbation theory. A systematic analysis of such enhancements, obtained from the aforementioned Coxeter-Dynkin diagram (Fig. \ref{fig:GDD}), was carried out in \cite{Fraiman:2023cpa}, showing that only eight of them are free of tachyons and that none of them minimize the 1-loop potential $\Lambda_\text{1-loop}$. Also interesting is the fact that all of these extrema have $\Lambda_\text{1-loop} > 0$. Implications for the existence of stable non-supersymmetric $AdS_3$ string vacua along the lines of \cite{Baykara:2022cwj} were also considered in this reference.

In great part, the control that we have over the theories just discussed comes from the high degree of supersymmetry in the heterotic string, which has 16 supercharges. In particular it constrains the spectrum to have only vector multiplets apart from the gravity multiplet, and correspondingly, the form of the moduli space is a coset space quotiented by a group of discrete symmetries. However, torus compactifications of the supersymmetric heterotic strings are not the only theories with 16 supercharges that we have at our disposal. As discovered by Chaudhury, Hockney and Lykken \cite{Chaudhuri:1995fk}, the fermionic formulation of the heterotic string allows to construct models with gauge symmetries of reduced rank and non-simply-laced type in four spacetime dimensions while preserving 16 supercharges. One of these possibilities, which reduces the rank by 8, can be realized in nine dimensions in a way that is transparent in the bosonic formulation \cite{Chaudhuri:1995bf} by means of the asymmetric orbifold construction of \cite{Narain:1986qm}. We will refer to this specific construction as the CHL string. It provides the heterotic description of the open string models with rank reduction discovered previously in \cite{Bianchi:1991eu}. See \cite{Font:2021uyw} (also \cite{Mikhailov:1998si}) for a detailed analysis. 

Both the Scherk-Schwarz mechanism and the bosonic formulation of the CHL string are realized by imposing a holonomy on the compactification circle. In the former case, this is the operation $(-1)^F$ with $F$ the spacetime fermion number, while in the latter it is the discrete gauge symmetry $R$ which exchanges the $E_8$ factors in the ten dimensional $E_8 \times E_8$ heterotic string. Quantum mechanically, this holonomy is realized by orbifolding the worldsheet theory by the corresponding aforementioned operations together with a half-shift $T$ along the circle direction. One can also consider implementing both operations $(-1)^F$ and $R$ as holonomies at the same time, a task that was carried out in detail recently in \cite{Nakajima:2023zsh}, where the spectrum was analyzed at various points in the moduli space and the behavior of $\Lambda_\text{1-loop}$ was studied at various asymptotic regions.   

We could ask if there are other inequivalent theories with broken supersymmetry and rank reduced by 8, perhaps exhibiting different behaviours with respect to their enhancing patterns, matter content and appearance of tachyons. In fact, we already know one such example, the $E_8$ string. This theory is obtained by orbifolding the $E_8 \times E_8$ string in ten dimensions by $(-1)^F R$; it has a tachyon which, unlike in the case of the rank 16 theories, cannot be removed by compactifying on a circle and turning on Wilson lines. One could also take the CHL string and compactify it further down to eight dimensions on a circle with $(-1)^F$ holonomy. This is a well defined non-supersymmetric theory with rank reduced by 8, but as it turns out, it does not correspond to a T-dual version of the aforementioned theory.  More precisely, their classical moduli spaces are not interconnected.   

\subsection{Classification proposal}

The scenario just presented poses a natural question: How many non-supersymmetric CHL-like heterotic strings are there? At this point we have three such theories. What we propose to do in this note is to make this question somewhat more well defined by introducing a class of non-supersymmetric heterotic strings which includes these three theories as particular examples. The defining property of such theories is that if one compactifies them on a suitable torus $T^n$ such that the target space is two-dimensional, the worldsheet theory always admits marginal deformations such that essentially it factorizes holomorphically (this notion will be made more precise promptly.) 

To justify this notion, we should recall that holomorphic factorization is already present in the supersymmetric heterotic string on $T^8$. Indeed, starting from either the $E_8\times E_8$ or the $SO(32)$ theories one can compactify on a $T^8$ without Wilson lines but with metric and B-field moduli such that the $T^8$ lattice takes the form $\Gamma_{8,8} \simeq E_8 \oplus E_8(-1)$ and so the Narain lattice becomes orthogonally split
\begin{equation}
	\Gamma_N \simeq (E_8\oplus E_8 \oplus E_8)\oplus E_8(-1) ~~~~~ \text{or} ~~~~~\Gamma_N \simeq (D_{16}^+ \oplus E_8)\oplus E_8(-1)\,.
\end{equation}
Here $D_{16}^+$ is the weight lattice of $Spin(32)/\mathbb{Z}_2$, which is even self-dual. Accordingly, the partition function is holomorphically factorized at these points, with the left-moving part a meromorphic\footnote{Following Schellekens \cite{Schellekens:1992db}, we take meromorphic to mean that there is only one character in the partition function. In the mathematics literature the associated objects are self-dual vertex operator algebras.} bosonic CFT with central charge $c = 24$ and the right moving part the sCFT with $c = 12$ based on the $E_8$ lattice. As we will review in the text, the same phenomenon is observed in the supersymmetric CHL string. More generally, the possible cyclic asymmetric orbifolds of the heterotic string with 16 supercharges in six dimensions seems to be in one-to-one correspondence with (families of) meromorphic bosonic CFTs through holomorphic factorization in two dimensions \cite{Fraiman:2022aik}. Another way of thinking about this relation is that the tensor product of meromorphic bosonic CFTs with $c = 24$ and the sCFT based on the $E_8$ lattice necessarily produce a heterotic worldsheet whose target space is two-dimensional and has 16 supercharges, so that if the rank of the gauge symmetry is non-zero, there will be massless scalars (see e.g. \cite{Harrison:2021gnp}). The claim in \cite{Fraiman:2022aik} is that such scalars always encode the compactification data of at least a $T^4$ without holonomies, hence the theory can always be decompactified to six or more dimensions without undoing its holonomies. In this way one fills out the list of all known theories of this type and produces more.  

One can alternatively construct a non-supersymmetric heterotic string in two dimensions by appropriately combining a meromorphic \textit{fermionic} CFT with $c = 24$ together with various spacetime characters and a right-moving contribution based on the $E_8$ lattice. As a prototypical example, consider the non-supersymmetric $O(16)\times O(16)$ heterotic string. We may procede inversely by repeating the compactification scheme for the supersymmetric case with $\Gamma_{8,8} \simeq E_8 \oplus E_8(-1)$. One finds that the partition function in 2D takes the form
\begin{equation}\label{nonsusyholo1}
	Z(\tau ,\bar \tau) = (Z_v \bar V_8 - Z_s \bar S_8 - Z_c \bar C_8 + Z_o \bar O_8)(\bar O_{16} + \bar S_{16})\,,
\end{equation}
where $Z_v$ is the partition function of the lattice $E_8 \oplus [D_8 \oplus D_8]^+$ whose root vectors produce the gauge bosons of the theory, and $Z_{c,s,o}$ are the partition functions associated to the remaining conjugacy classes producing fermions and scalars (we follow the conventions and rely on formulae nicely summarized in Appendix A of \cite{Nakajima:2023zsh}). The factor $\bar O_{16} + \bar S_{16}$ is the partition function of the right moving compact bosons moving on the $E_8$ lattice. The dependence on the modular parameters is implied, and bars denote dependence on $\bar \tau$. The partition function \eqref{nonsusyholo1} is factorized in the sense that it can be written as the vector product 
\begin{equation}
	Z(\tau, \bar \tau) = \vec Z_L(\tau) \cdot \vec Z_R(\bar \tau)\,,                                 
\end{equation}
with
\begin{equation}
	\vec Z_L(\tau) = (Z_v, Z_s, Z_c, Z_o)\,, ~~~~~ \vec Z_R(\bar \tau) = (\bar V_8, -\bar S_8, -\bar C_8, \bar O_8) (\bar O_{16} + \bar S_{16})\,.
\end{equation}
The vector valued character $\vec Z_L$ then defines a fermionic CFT with $c = 24$ and rank also 24. It can then be shown that every other CFT of this type, of which there are 273, can be realized analogously in the same theory through marginal deformations. It then follows that starting from any of these 273 CFTs one obtains a heterotic worldsheet which ultimately decompactifies to any of the rank 16 non-supersymmetric heterotic strings in ten dimensions.\footnote{This phenomenon of splitting into different theories upon decompactification is unique to this example. For the asymmetric orbifolds we consider, compactification and decompactification on circles (without holonomies) is a one-to-one operation.} 

What we wish to do in this note is to provide evidence that the classification scheme described above for supersymmetric heterotic strings extends to the non-supersymmetric case. Namely that if we start from the classification of meromorphic fermionic CFTs with $c = 24$ we can construct a family of non-supersymmetric heterotic theories in two dimensions which can be decompactified to six or more dimensions, including those three theories with rank reduced by 8 considered earlier. Fortunately for us, this classification was obtained recently by Höhn and Möller \cite{Hohn:2023auw}. The methods consist in a procedure which in physical terms can be thought of as fermionization, as explained in detail in \cite{BoyleSmith:2023xkd} for theories with central charge 16 (see also \cite{Rayhaun:2023pgc} for an alternative classification method with results up to $c = 22 \nicefrac{1}{2}$.) Among our tasks will be the identification of various string compactifications with the families of CFTs listed in the results of \cite{Hohn:2023auw}. In this note we will restrict to a detailed analysis of the case of rank reduction by 8, while for the remaining cases we will give various arguments in favour of the validity of the relation.

Along the way we will make some comments regarding the relation of non-supersymmetric heterotic strings and the so-called MSDS vacua of \cite{Kounnas:2008ft}, which appear at special factorization points in the moduli space of compactifications to two dimensions where the right-moving CFT is not the sCFT based on the $E_8$ lattice but rather the CFT constructed with 24 real fermions with current algebra $8\,\mathfrak{su}{2,2}$. The partition function of this right-moving CFT is numerically equal to 24, and so the number of level-matched bosons and fermions is exactly the same at every massive level. 

Part of the motivation behind this paper was to understand the mechanism by which gauge group ranks are reduced in different string theories and the phenomena to which they are connected. In the supersymmetric case, the perturbative constructions of the type considered here are dual to F/M-theory on K3 surfaces with frozen singularities \cite{Witten:1997bs,deBoer:2001wca}, thus the non-supersymmetric version explored here could in some way retain a relationship with a geometric picture, providing insight into the non-supersymmetric phase of F/M-theory. In this spirit we also mention that such a relationship was studied in the context of so-called Nikulin orbifolds \cite{Acharya:2022shu}, which however exhibit a more complicated configuration space similar to four dimensional $\mathcal{N} = 2$ theories. Another context in which CHL strings seem to be relevant to non-supersymmetric theories was explored in \cite{Coudarchet:2021qwc} through brane supersymmetry breaking in its orientifold dual which is devoid of a tree level NSNS tapole (see \cite{Mourad:2017rrl} for a recent review on brane supersymmetry breaking). Ultimately our goal is to put in good order as much as we can our understanding of the non-supersymmetric phase of string theory. Along these lines see also \cite{Basile:2023knk} for a recent study of potential anomalies in the ten dimensional non-supersymmetric heterotic strings. 

This paper is organized as follows. In Section \ref{s:susy} we review the relationship between supersymmetric heterotic strings and chiral bosonic CFTs with $c = 24$, proving it explicitly for the Narain and CHL string through partition function manipulations. We then generalize these considerations to non-supersymmetric strings in Section \ref{s:nonsusy}, going carefully through various constructions with rank reduced by 8, apart from the full rank case. We show in two cases that two different theories are T-dual in line with our overall proposal. At the end of this section we give some comments on the behaviour of tachyons in the classical moduli spaces of these theories and leave open an interesting puzzle for another kind of non-supersymmetric orbifold. In Section \ref{s:remaining} we go over the remaining fermionic CFT genera and discuss which ones are straightforwardly related to non-supersymmetric strings in six dimensions and which ones are not. We give some concluding remarks in Section \ref{s:conc}.

\section{The supersymmetric case}
\label{s:susy}
We start by reviewing the relation between supersymmetric heterotic strings and bosonic meromorphic CFTs with $c = 24$. This connection was explored in detail in \cite{Fraiman:2022aik} by exploiting various properties of the underlying charge lattices\footnote{By charge lattice we mean the Narain lattice or its analogs for rank reduced theories, i.e. the lattice whose automorphisms define the T-duality group of the theory. In two dimensions this lattice can split orthogonally into two charge lattices, each of which corresponds to some (anti)holomorphic CFT.} which specify the global structure of the moduli spaces. Here we will prove the relation explicitly for the standard heterotic strings (see also \cite{Kachru:2016ttg}) and the CHL string by manipulating their partition functions and matching them with those of the corresponding meromorphic CFTs.

\subsection{The standard Narain component}
The bosonic CFTs in Schellekens' list can be arranged into various families or genera using as a criterion the structure of the underlying electric charge lattice \cite{Hohn:2017dsm}. The first genus, labeled $\mathcal{A}$, contains the 24 meromorphic CFTs based on the Niemeier lattices $N_I$. Since these lattices are even and self-dual, it follows that
\begin{equation}
	\Gamma_{24,8} \simeq N_I \oplus E_8(-1)\,,
\end{equation} 
which implies that the Narain lattice $\Gamma_{24,8}$ splits orthogonally into two Euclidean lattices at 24 special points in its moduli space $O(24,8,\mathbb{R})/O(24,\mathbb{R})\times O(8,\mathbb{R})$. For two of these points, $N_I = 3\, E_8\,,  D_{16}^+\oplus E_8$, corresponding to the two holomorphic factorization points discussed in section \ref{s:intro} for the heterotic strings on $T^8$ without Wilson lines. Concretely, the partition functions read
\begin{equation}
	Z(\tau,\bar \tau) = Z_L(\tau) \times Z_R(\bar \tau)
\end{equation}
with $Z_L = (O_{16}+S_{16})^3,~ (O_{32}+S_{32})(O_{16}+S_{16})$, respectively, and $Z_R = (\bar O_{16} + \bar S_{16})(\bar V_8 - \bar S_8)$. As before, dependence on $\tau$ is implied for holomorphic quantities and the bar denotes antiholomorphicity, and $V_n, C_n, S_n, O_n$ are $SO(n)$ characters. By also turning on the Wilson lines, one can reach the remaining 22 points in moduli space. This establishes a correspondence between the ten-dimensional supersymmetric heterotic strings and the 24 meromorphic CFTs in genus $\mathcal{A}$.

\subsection{The CHL string}

The scenario above extends to theories with rank reduction. For the CHL string, the charge lattice in nine dimensions is $\Gamma_{1,1}\oplus E_8$ \cite{Mikhailov:1998si}. Compactifying on $T^7$, this lattice is extended to $\Gamma_{7,7}(2)\oplus \Gamma_{1,1}\oplus E_8$, with the number in parentheses denoting a scaling by $2$ of the quadratic form of $\Gamma_{7,7}$. One can then show that this lattice is isomorphic to $E_8 \oplus D_8^*(2)\oplus E_8(-2)$, where $D_8^*$ is the lattice dual to $D_8$, suggesting that there is a corresponding holomorphic factorization of the partition function \cite{Fraiman:2022aik}. Let us check this example in detail. 
\subsubsection{CHL partition function}
The partition function for the CHL string with vanishing Wilson lines in nine dimensions can be written as follows. Consider first the partition function for the $E_8 \times E_8$ heterotic string on a circle, 
\begin{equation}
	\begin{split}
		Z(\tau,\bar \tau) &= \frac{1}{\tau_2^{7/2}\eta^{7}\bar \eta^7} \left(\frac{1}{\bar \eta \eta}\sum_{(p_R;p_L) \in \Gamma_{1,1}} \bar q^{\tfrac12 p_R^2} q^{\tfrac12 p_L^2} \right)\frac{1}{\eta^{16}}\Theta_8^2 (\bar V_8 - \bar S_8)\,,\\
		&\equiv \frac{1}{\tau_2^{7/2}\eta^{7}\bar \eta^7} Z_\Lambda (\bar V_8 - \bar S_8)\,,
	\end{split}
\end{equation}
where $\eta$ is the Dedekind eta function, $\Theta_8 = \sum_{p_L \in E_8}q^{p_L^2/2}$ is the theta function of the $E_8$ lattice; in particular, $\Theta_8 = \eta^8(O_{16}+S_{16})$. Under the Chaudhuri-Polchinski orbifold $g = RT$, only the lattice contribution $Z_\Lambda$ is affected:
\begin{equation}
	Z_\Lambda \overset{g}{\to} \frac12 \left(Z_\Lambda^{1,1}+ Z_\Lambda^{1,g} + Z_\Lambda^{g,1} + Z_\Lambda^{g,g} \right)\,,
\end{equation}
with \cite{Mikhailov:1998si,Font:2021uyw}
\begin{align}\label{Zchl1}
	Z_\Lambda^{1,1} &= ~Z_\Lambda\,,\\
	Z_\Lambda^{1,g} &= ~\frac{1}{\bar\eta \eta^{17}}\left(\frac{2\eta^3}{\vartheta_2}\right)^4 \sum_{(p_R;p_L) \in \Gamma_{1,1}} \bar q^{\tfrac12 p_R^2} q^{\tfrac12 p_L^2} e^{i\pi n} \Theta_8(2\tau)\,,\label{Zchl2}\\
	Z_\Lambda^{g,1} &= ~\frac{1}{\bar\eta \eta^{17}}\left(\frac{\eta^3}{\vartheta_4}\right)^4 \sum_{(p_R;p_L) \in \Gamma_{1,1}'} \bar q^{\tfrac12 p_R^2} q^{\tfrac12 p_L^2}\Theta_8(\tau/2)\,,\label{Zchl3}\\
	Z_\Lambda^{g,g} &= -\frac{1}{\bar\eta \eta^{17}}\left(\frac{\eta^3}{\vartheta_3}\right)^4 \sum_{(p_R;p_L) \in \Gamma_{1,1}'} \bar q^{\tfrac12 p_R^2} q^{\tfrac12 p_L^2}e^{i \pi n}\Theta_8'(\tau/2)\,,\label{Zchl4}
\end{align}
where $\vartheta_i$ are the usual Jacobi theta functions, and we have defined the alternating sum
\begin{equation}\label{theta8'}
	\Theta_8'(\tau) \equiv \sum_{p \in E_8}q^{p^2/2} (-1)^{p^2}\,.
\end{equation}
Note the dependences on the scaled modular parameters $2\tau$ and $\tau/2$. $\Gamma_{1,1}'$ is defined as the lattice $\Gamma_{1,1}$ shifted by $\delta$ such that the winding numbers $w$ are half-integral. $n$ is the quantized momentum along the circle.
\subsubsection{Compactifying and factorizing}
Now we must compactify the theory on $T^7$ and do marginal deformations so as to factorize these four partition functions. The first operation transforms
\begin{equation}
	\frac{1}{\tau_2^{7/2}\bar \eta^7\eta^7} \to \frac{1}{\bar \eta^7 \eta^7}\sum_{(p_R;p_L) \in \Gamma_{7,7}} \bar q^{\tfrac12 p_R^2} q^{\tfrac12 p_L^2}\,,
\end{equation}
and together with the $\Gamma_{1,1}$ contribution allows to easily factorize $Z_\Lambda^{1,1}$ by deforming
\begin{equation}\label{deform}
	\frac{1}{\bar \eta^8 \eta^8}\sum_{(p_R;p_L) \in \Gamma_{8,8}} \bar q^{\tfrac12 p_R^2} q^{\tfrac12 p_L^2} \to \frac{1}{\bar \eta^8 \eta^8} \bar \Theta_8 \Theta_8 = (\bar O_{16} + \bar S_{16})(O_{16} + S_{16})
\end{equation}
just as for the $E_8 \times E_8$ heterotic string on $T^8$. We have then
\begin{equation}
	Z^{1,1} \to (O_{16} + S_{16})^3(\bar O_{16} + \bar S_{16})(\bar V_8 - \bar S_8) \equiv Z_{L}^{1,1}(\tau) \times {Z}_R(\bar \tau)\,.
\end{equation}

In $Z_\Lambda^{1,g}$, the phase $e^{i \pi n}$ splits $\Gamma_{1,1}$ into the index 2 sublattice with even momenta and its complement. Under the marginal deformation \eqref{deform}, the shift vector $\delta$ is mapped to $(1,0^7) \in E_8 \otimes \mathbb{R}$ and the corresponding phase splits the lattice $E_8$ into $D_8 \subset E_8$ and its complement given by the the spinor class of $SO(16)$, hence
\begin{equation}\label{deform2}
	\frac{1}{\bar \eta^8 \eta^8}\sum_{(p_R;p_L) \in \Gamma_{8,8}} \bar q^{\tfrac12 p_R^2} q^{\tfrac12 p_L^2}e^{i\pi n} \to (\bar O_{16} + \bar S_{16})(O_{16} - S_{16})\,,
\end{equation}
and so $Z^{1,g}$ also factorizes holomorphically into $Z^{1,g}_L(\tau) \times Z_R(\bar \tau)$ with
\begin{equation}
	Z^{1,g}_L = \frac{1}{\eta^{16}}\left(\frac{2\eta^3}{\vartheta_2}\right)^4 (O_{16} - S_{16}) \Theta_8(2\tau)\,.
\end{equation}
Following the same reasoning for $Z_\Lambda^{g,1}$ and $Z_\Lambda^{g,g}$ or alternatively applying the modular transformations $S$ and $T$ sequentially on $Z^{1,g}_L$, we obtain
\begin{equation}
	\begin{split}
		Z^{1,g}_L &= \frac{1}{ \eta^{16}}\left(\frac{\eta^3}{\vartheta_4}\right)^4 (V_{16} + C_{16})\Theta_8(\tau/2)\,,\\
		Z^{g,g}_L &= \frac{1}{ \eta^{16}}\left(\frac{\eta^3}{\vartheta_3}\right)^4 (V_{16} - C_{16})\Theta_8'(\tau/2)\,.
	\end{split}
\end{equation}
The full partition function is thus holomorphically factorized with $Z(\tau,\bar \tau) = Z_L(\tau) \times Z_R(\bar \tau)$. 

Doing a $q$-expansion (with $q = e^{2\pi i \tau}$) we find that 
\begin{equation}\label{twistL}
	\begin{split}
		Z_L^{1,1} &= q^{-1} + 744 + 196884 q + O(q^2)\,, ~~~~~~~~~~
		Z_L^{1,g} = q^{-1} - 8 + 276 q + O(q^2)\,,\\
		2^{-4}~Z_L^{g,1} &= 1 + 256q^{1/2} + 6144q + O(q^{3/2})\,, ~~~
		2^{-4}~Z_L^{g,g} = 1 - 256q^{1/2} + 6144q + O(q^{3/2})\,,	 
	\end{split}
\end{equation}
and so
\begin{equation}\label{chlqexp}
	Z_L \equiv \frac12 (Z_L^{1,1}+Z_L^{1,g}+Z_L^{g,1}+Z_L^{g,g}) = q^{-1} + 384 + 196884 q + O(q^2)\,.
\end{equation}
We then identify this partition function with that of the meromorphic bosonic CFT with current algebra $\mathfrak{e}_{8,2}\oplus \mathfrak{so}_{17,1}$ which has dimension $248 + 136 = 384$ (item 62 in Schellekens' list \cite{Schellekens:1992db}). This is in accordance with the data extracted from the orthogonally split charge lattice $E_8 \oplus D_8^*(2)\oplus E_8(-2)$, where the positive definite part $E_8 + D_8^*(2)$ has elements furnishing the short and long roots of the algebra $\mathfrak{e}_{8,2}\oplus \mathfrak{so}_{17,1}$ while the negative definite part $E_8(-2)$ corresponds to the right-moving CFT states. 

\subsubsection{Relation to the classification}

This rank 16 CFT along with 16 others lies in genus $\mathcal{B}$ in the classification of \cite{Hohn:2017dsm}. The underlying rank 16 lattices for these theories all lie in the same lattice genus, which implies that the CHL charge lattice can be orthogonally split such that the Euclidean part corresponds to any one of them. This means that genus $\mathcal{B}$ is in one-to-one correspondence with the CHL string. 

This result may be reinterpretted as telling us that the CHL orbifold can be realized holomorphically in two dimensions, or equivalently that all the meromorphic bosonic CFTs with $c = 24$ and rank 16 can be obtained from those of rank 24 by orbifolding. This latter result is known to hold for every item in Schellekens list \cite{Hohn:2020xfe}. What we have shown here is that for rank 16, the corresponding orbifold is nothing but the one of Chaudhuri-Polchinski \cite{Chaudhuri:1995bf} in some T-dual frame. That this extends to every other possibility, namely the genera $\mathcal{C},...,\mathcal{K}$,  is one of the central claims of \cite{Fraiman:2022aik}. Explicit realizations of the 6D heterotic theories corresponding to the genera $\mathcal{D}, \mathcal{J}$ and $\mathcal{K}$ are still under investigation. For the remaining cases, the constructions are known \cite{deBoer:2001wca} and the above analysis should extend in a straightforward manner.  

\section{Non-supersymmetric strings}\label{s:nonsusy}
Let us now consider the case of non-supersymmetric strings and their relation to meromorphic fermionic CFTs with $c = 24$, which have been classified recently in \cite{Hohn:2023auw}. They are obtained from their bosonic counterparts discussed in section \ref{s:susy}, and are similarly arranged into various families or genera. We will focus our attention on the genera $\mathcal{A}_I$, $\mathcal{B}_I$, $\mathcal{B}_{IIa}$, $\mathcal{B}_{IIb}$ and $\mathcal{B}_{III}$, comprising all fermionic CFTs with rank 24 and 16 as indicated by the big letters $\mathcal{A}$ and $\mathcal{B}$. The subscript denotes how the theories are obtained from bosonic ones, and reflects the properties of their spectra. In what follows we will simply match the genera to specific string compactifications, and in the process we will come back to the meaning of these subscripts (cf. subsection \ref{ss:nonsusychl}). At this point we note that the existence of four different genera with rank 16 suggest the existence of four non-supersymmetric CHL strings; this expectation will be confirmed in the following.

The torus partition functions of all the fermionic CFTs in \cite{Hohn:2023auw} are specified by a vector-valued character
\begin{equation}
	\vec Z_L(\tau) = (Z_v,Z_s,Z_c,Z_0)\,,
\end{equation}
where the notation reflects what we will be using them for. Concretely, these CFTs give rise to heterotic worldsheets with partition function (c.f. section \ref{s:intro})
\begin{equation}\label{factorization}
	\vec Z_L(\tau) \cdot \vec Z_R(\bar \tau),
\end{equation}
where 
\begin{equation}
	\vec Z_R(\bar \tau) = (\bar V_8, -\bar S_8, -\bar C_8, \bar O_8) (\bar O_{16} + \bar S_{16})\,.
\end{equation}
From \cite{Hohn:2023auw}, $\vec Z_L$ is specified as a $q$-expansion by the integers $a,b,\ell$ in
\begin{equation}\label{Zt}
	\vec Z_L = F + a ~(1,1,0,0) + b ~ (1,0,1,0) + \ell ~G\,,
\end{equation}
where
\begin{equation}
	F = 
	\begin{pmatrix}
		q^{-1} - 24 + 98580q + O(q^2)\\
		98304q + O(q^2)\\
		98304q + O(q^2)\\
		4096q^{1/2} + O(q^{3/2})\\
	\end{pmatrix}\,, ~~~~~
	G = 
	\begin{pmatrix}
		- 24 + 2048q + O(q^2)\\
		-24 - 2048 + O(q^2)\\
		-24 - 2048 + O(q^2)\\
		q^{-1/2} + 276q^{1/2} + O(q^{3/2})\\
	\end{pmatrix}\,.
\end{equation}
The exact expressions for $F$ and $G$ can be found in \cite{Hohn:2023auw}, but these expansions will suffice for the purpose of matching the various $\vec Z_L$'s.

\subsection{Non-supersymmetric rank 16 heterotic strings}\label{ss:rk16}
As a warm up let us start by explaining in more detail the $T^8$ compactification of the rank 16 non-supersymmetric heterotic strings briefly discussed in section \ref{s:intro}. For concreteness let us take the ten dimensional $O(16)\times O(16)$ heterotic string. Its partition function reads
\begin{equation}\label{o16z}
	Z(\tau,\bar \tau) = \frac{1}{\tau_2^4 \bar \eta^8 \eta^8} \times \frac{1}{\eta^{16}}\left(Z_v \bar V_8 - Z_s \bar S_8 - Z_c \bar C_8 + Z_o \bar O_8   \right)\,,
\end{equation}
where $Z_{v,s,c,o}$ can be conveniently written as 
\begin{equation}\label{Zclasses}
	\begin{split}
		\frac{1}{\eta^{16}}Z_v &= O_{16}^2 + S_{16}^2\,, ~~~~~ ~~~~~ \frac{1}{\eta^{16}}Z_s = 2 O_{16}S_{16}\,,\\
		\frac{1}{\eta^{16}}Z_c &= V_{16}^2 + C_{16}^2\,, ~~~~~ ~~~~~ \frac{1}{\eta^{16}}Z_o = 2 V_{16}C_{16}\,. 
	\end{split}
\end{equation}
As for any of the ten-dimensional theories, this partition function is already factorized as in eq. \eqref{factorization}. Compactifying on a $T^8$ with Narain lattice $\Gamma_{8,8} = E_8\oplus E_8(-1)$ preserves this factorization, with
\begin{equation}
	\vec Z_L^t = (O_{16}+S_{16})
	\begin{pmatrix}
		O_{16}^2 + S_{16}^2\\2 O_{16}S_{16}\\V_{16}^2 + C_{16}^2\\2 V_{16}C_{16}
	\end{pmatrix} = 
	\begin{pmatrix}
	q^{-1} + 488 + 98580q + O(q^2)\\
	256 + 98304q + O(q^2)\\
	256 + 98304q + O(q^2)\\
	4096q^{1/2} + O(q^{3/2})
\end{pmatrix}\,.
\end{equation}	
This $q$-expansion corresponds to eq. \eqref{Zt} with $a = b = 256$ and $\ell = 0$ (cf. entry 14 in Table 12 of \cite{Hohn:2023auw}), which sits in the genus $\mathcal{A}_I$. 

There are 272 other elements in genus $\mathcal{A}_I$, which can be reached by doing marginal deformations on the heterotic worldsheet above. To see this, consider the charge lattice of the theory\cite{Fraiman:2023cpa},
\begin{equation}
	\Upsilon_{24,8} \simeq \left(\Gamma_{8,8}\oplus [2\,D_8]^+\right)^*,
\end{equation}  
where $([2\,D_8]^+)^*$ is the charge lattice of the ten dimensional $O(16)\times O(16)$ string. The moduli space of the theory corresponds to boosts of this lattice, and we may as well work with its dual $\Upsilon_{24,8}^*$ which corresponds to the sector furnishing spacetime vectors. For the moduli chosen above, this lattice splits orthogonally as
\begin{equation}
	\Upsilon_{24,8}^* \simeq E_8(-1)\oplus E_8 \oplus [2\, D_8]^+\,, 
\end{equation}
and any other splitting with an $E_8(-1)$ part will correspond to a factorization of the partition function. It follows that the lattice genus containing the lattice $E_8\oplus[2\, D_8]^+$ gives all such factorizations, since $\Upsilon_{24,8}^* \simeq E_8(-1)\oplus L$ for $L$ in this genus. There are 273 lattices in this genus and they in turn correspond to the CFT genus $\mathcal{A}_\text{I}$. It should be noted that this lattice construction underlies the classification of \cite{Hohn:2023auw}; the important statement here is that the charge lattice of the full worldsheet can be boosted to produce the genus. Torus compactification of the rank 16 non-supersymmetric heterotic string theories is therefore in one-to-one correspondence with genus $\mathcal{A}_\text{I}$.

\subsection{The $E_8$ string}
Let us now consider the remaining ten dimensional non-supersymmetric heterotic string, namely the $E_8$ string. This theory is obtained from the $E_8 \times E_8$ heterotic string through an orbifold with $g = R(-1)^F$, giving rise to the partition functions
\begin{align}\label{ZE8}
	Z^{1,1} &=\frac{1}{\tau_2^4 \bar \eta^8 \eta^8} \times  \frac{1}{\eta^{16}} \Theta_8^2(\bar V_8 - \bar S_8)\,,\\
	Z^{1,g} &= \frac{1}{\tau_2^4 \bar \eta^8 \eta^8} \times \frac{1}{\eta^{16}} \left(\frac{2\eta^3}{\vartheta_2}\right)^4 \Theta_8(2\tau) (\bar V_8 + \bar S_8)\,,\\
	Z^{g,1} &= \frac{1}{\tau_2^4 \bar \eta^8 \eta^8} \times \frac{1}{\eta^{16}} \left(\frac{\eta^3}{\vartheta_4}\right)^4 \Theta_8(\tau/2)(\bar O_8 - \bar C_8)\,,\\
	Z^{g,g} &=\frac{1}{\tau_2^4 \bar \eta^8 \eta^8} \times  \frac{1}{\eta^{16}} \left(\frac{\eta^3}{\vartheta_3}\right)^4 \Theta_8'(\tau/2)(\bar O_8 + \bar C_8)\,.
\end{align}
Compactifying on a torus $T^8$ with orthogonally split Narain lattice as before, and combining appropriately the different terms, we obtain $\vec Z_L$ with $q$-expansion
\begin{equation}\label{E8ZL}
	\vec Z_L^t = 
	\begin{pmatrix}
		q^{-1} + 496 + 100628 q + O(q^2)\\
		248 + 96256q + O(q^2)\\
		248 + 96256q + O(q^2)\\
		q^{-1/2} + 4372q^{1/2} + O(q^{3/2})
	\end{pmatrix}\,.
\end{equation}
This corresponds to \eqref{Zt} with $a = b = 272$ and $\ell = 1$ (cf. entry 593 in Table 12 of \cite{Hohn:2023auw}). The fermionic CFT is in genus $\mathcal{B}_\text{III}$, which contains in total 24 items. 

The charge lattice of this theory is simply $\Gamma_{8,8}(2) \oplus E_8 $, and the factorization above corresponds to the lattice polarization
\begin{equation}
	\Gamma_{8,8}(2) \oplus E_8 \simeq E_8(-2)\oplus E_8(2)\oplus E_8\,.
\end{equation}
Just as for the CHL string, the lattice $E_8(2)\oplus E_8$ is associated to the gauge algebra $\mathfrak{e}_8+\mathfrak{e}_{8,2}$, which is reflected in the numerical term $496$ in the expression \eqref{E8ZL}. Similarly to the case of gneus $\mathcal{A}_\text{I}$, the lattice genus containing $E_8(2)\oplus E_8$ has the 24 lattices corresponding to the CFTs in genus $\mathcal{B}_\text{III}$. Note however that the charge lattice in this theory is unlike that of the previous case. It does not split into conjugacy classes associated to the spacetime Lorentz group representations. This is reflected on the fact that in ten dimensions the massless spinors and co-spinors sit in the adjoint representation just as for the gauge bosons. 

\subsection{Non-supersymmetric CHL string} \label{ss:nonsusychl}
Let us move on to the simplest non-trivial case, corresponding to the non-supersymmetric version of the CHL string realized as an orbifold with $g = RT(-1)^F$ as discussed in section \ref{s:intro}. The effects of including the operator $(-1)^F$ lead to a partition function which can be easily written in terms of the building blocks $Z^{1,1}_\Lambda,...,Z^{g,g}_\Lambda$ for the CHL string (eqs. \eqref{Zchl1} to \eqref{Zchl4}) \cite{Nakajima:2023zsh},
\begin{equation}
	\begin{split}
		Z(\tau,\bar \tau) = \frac{1}{\tau^{7/2}\bar \eta^7 \eta^7} [(Z_\Lambda^{1,1}+Z_\Lambda^{1,g})\bar V_8 &- (Z_\Lambda^{1,1}-Z_\Lambda^{1,g})\bar S_8 \\ &~~~~~~~~~~-(Z_\Lambda^{g,1}+Z_\Lambda^{g,g})\bar C_8 + (Z_\Lambda^{g,1}-Z_\Lambda^{g,g})\bar O_8]\,. 
	\end{split}
\end{equation} 
After compactifying on $T^7$, each term can be factorized just as for the CHL string. Using the $q$-expansions in eq. \eqref{chlqexp}, we obtain
\begin{equation}
	\vec Z_L^t = 
	\begin{pmatrix}
		Z_L^{1,1} + Z_L^{1,g}\\
		Z_L^{1,1} - Z_L^{1,g}\\
		Z_L^{g,1} + Z_L^{g,g}\\
		Z_L^{1,g} - Z_L^{g,g}
	\end{pmatrix} = 
	\begin{pmatrix}
		q^{-1} + 368 + 98580 q + O(q^2)\\
		376 + 98304q + O(q^2)\\
		16 + 98304q + O(q^2)\\
		4096q^{1/2} + O(q^{3/2})
	\end{pmatrix}\,,
\end{equation}
where $t$ means transpose. This expression corresponds to \eqref{Zt} with $a = 376$, $b = 16$ and $\ell = 0$, which is  the entry 517 in Table 12 of \cite{Hohn:2023auw}. It follows that the heterotic orbifold above corresponds to genus $\mathcal{B}_\text{IIb}$ under the assumption that all the corresponding worldsheets formed with the rest of the CFTs therein are connected through marginal deformations. Unlike the cases examined before, it is not straightforward to prove this latter claim for this theory. 

To proceed, we come back to the question of how the various fermionic CFTs are obtained from the bosonic ones. As discussed in \cite{Hohn:2023auw}, genera with subscripts $I$, $IIa$ and $IIb$ are obtained by using shift vectors in the lattice of the parent theory with different norms. Torus compactifications of the supersymmetric heterotic strings, corresponding to genus $\mathcal{A}$, have charge lattices with only one kind of shift vector up to automorphisms, which in a specific frame are associated to half-shifts on a compact cycle and so to the usual Scherk-Schwarz reductions (as explained in the Introduction). The associated fermionic CFT genus is $\mathcal{A}_I$, and as we will see in section \ref{ss:SSCHL}, the Scherk-Schwarz reduction of the CHL string produces in turn the genus $\mathcal{B}_I$. This means that in the  CHL string, the shift associated to the subscript $I$ lies along a factor $\Gamma_{1,1}(2)$ in the Mikhailov lattice, and so can only be constructed in eight or less dimensions, where this lattice takes the form $\Gamma_{1,1}(2) \oplus \cdots \oplus \Gamma_{1,1}(2) \oplus \Gamma_{1,9}$. In nine dimensions, however, we can choose shifts lying on $\Gamma_{1,9}$. The obvious choice is to take a half-shift along the circle, i.e. a shift located in $\Gamma_{1,1}$, producing the theory under consideration which we are associating to genus $\mathcal{B}_{IIb}$. As we will see below in section \ref{ss:CPE7}, there is another inequivalent shift producing a nine-dimensional theory corresponding to genus $\mathcal{B}_{IIa}$.

The idea is that the CHL string supports, up to T-duality, three inequivalent shifts, each of which can be combined with the operation $(-1)^F R$ to obtain a modular invariant theory. Together with the fact that the CHL string factorizes holomorphically in two dimensions yielding every bosonic CFT with rank 16 and the fact that these bosonic CFTs yield the fermionic CFTs in the three genera under discussion, this shows the validity of the connectedness claim above; the heterotic worldsheets constructed with each genera lie in a connected classical moduli space component. Genera of type $III$ do not involve shifts in their construction, and we have already proven above that for the associated theory, the $E_8$ string, the connectedness claim holds.  

\subsection{$\mathbb{Z}_2$ orbifold of the $O(16)\times O(16)$ string}\label{ss:CPO16}
Before moving onto the next inequivalent theory, we will consider another non-supersymmetric construction with rank reduction which is natural to think of. Just as the $E_8 \times E_8$ string, the non-supersymmetric $O(16)\times O(16)$  string has a $\mathbb{Z}_2$ symmetry $R$ exchanging the two gauge factors.\footnote{We note that the $\mathbb{Z}_2$ orbifold $R(-1)^F$ can be carried out in ten dimensions, but for this theory as well as for the $E_7 \times SU(2)\times E_7 \times SU(2)$ theory one does not obtain anything new; the result is the $E_8$ string again \cite{Forgacs:1988iw}. Making the orbifold freely acting leads instead to inequivalent theories.} It is then possible to perform a CHL-like orbifold starting from this theory compactified on a circle. We will carry out this computation and demonstrate that it is T-dual to the one examined in the previous subsection. In other words, we will show that it corresponds to the CFT genus $\mathcal{B}_{IIb}$. We then make explicit contact with this genus using the compactification and factorization scheme already employed, reinforcing the T-duality claim and the connectedness claim elaborated on previously.

We aim thus to compute the partition function of the orbifold of the $O(16)\times O(16)$ heterotic string on $S^1$ with $g = RT$, where $R$ exchanges the two $O(16)$ factors and $T$ is a half-shift along the circle. To make things more readable, we define the following functions:
\begin{equation}
	\zeta_2(\tau)  = \left(\frac{2\eta^3(\tau)}{\vartheta_2(\tau)}\right)^4\,, ~~~~~ \zeta_3(\tau)  = \left(\frac{\eta^3(\tau)}{\vartheta_4(\tau)}\right)^4\,, ~~~~~ \zeta_4(\tau)  = -\left(\frac{\eta^3(\tau)}{\vartheta_3(\tau)}\right)^4\,,
\end{equation}
where the numbering is according to appearance in the orbifolding procedure.  We also define
\begin{equation}\label{defS}
	\begin{split}
		Z_{S^1}^{(1)}(\tau) &= \sum_{(p_R;p_L) \in \Gamma_{1,1}} \bar q^{\tfrac12 p_R^2} q^{\tfrac12 p_L^2}\,, ~~~~~ Z_{S^1}^{(2)}(\tau) = \sum_{(p_R;p_L) \in \Gamma_{1,1}} \bar q^{\tfrac12 p_R^2} q^{\tfrac12 p_L^2}e^{i\pi n}\,,\\
		Z_{S^1}^{(3)}(\tau) &= \sum_{(p_R;p_L) \in \Gamma_{1,1}'} \bar q^{\tfrac12 p_R^2} q^{\tfrac12 p_L^2}\,, ~~~~~ Z_{S^1}^{(4)}(\tau) = \sum_{(p_R;p_L) \in \Gamma_{1,1}'} \bar q^{\tfrac12 p_R^2} q^{\tfrac12 p_L^2}e^{i\pi n}\,,
	\end{split}
\end{equation}
where again $n$ is the quantized momentum along the circle and $\Gamma_{1,1}'$ denotes the lattice $\Gamma_{1,1}$ shifted such that the winding number takes half-integral values.

\subsubsection{Orbifolding procedure}
The partition function of the parent theory reads (cf. eq. \eqref{o16z})
\begin{equation}\label{appZ}
	Z(\tau,\bar \tau) = \frac{1}{\tau_2^{7/2} \bar \eta^7 \eta^7}  \times \frac{1}{\eta^{16}}\left(Z_v \bar V_8 - Z_s \bar S_8 - Z_c \bar C_8 + Z_o \bar O_8   \right)\,,
\end{equation}
where, in distinction to eq. \eqref{o16z}, the blocks $Z_{v,s,c,o}$ have a circle contribution $(\bar \eta \eta)^{-1}Z_{S^1}^{(1)}$. We will proceed by effectively applying the asymmetric orbifold procedure to each of the four terms in $Z(\tau,\bar \tau)$. A detailed exposition of how this is done in the case of the supersymmetric CHL string is presented in Appendix A of \cite{Font:2021uyw}, but certain considerations make it so that the full details are not needed to get the desired results here. 

Concretely, let us write the full partition function resulting from operating on $Z_v$:
\begin{equation}
	Z_v \to \frac12 (Z_v^{1,1} + Z_v^{1,g} + Z_v^{g,1} + Z_v^{g,g})\,.
\end{equation}
Because of the outer automorphism $R$, the term $Z_{v}^{1,g}$ is given by a sum over the invariant lattice inside the vector class $\Gamma_{1,1}+[2\, D_8]^+$. Because of the spinor element $(\tfrac12^{16})$, which is invariant under $R$, the full invariant lattice is $\Gamma_{1,1}\oplus E_8(2)$ rather than $\Gamma_{1,1}\oplus D_8(2)$. This is the same invariant lattice as for the supersymmetric CHL string, and so $Z_v^{1,g}$ takes exactly the same form as $Z_\Lambda^{1,g}$ in eq. \eqref{Zchl2} if one includes the $\eta^{-{16}}$ prefactor. Moreover, since $Z_v^{g,1}$ and $Z_v^{g,g}$ are obtained by applying modular transformations on $Z_v^{1,g}$, they are also just as for the supersymmetric case (cf. eqs. \eqref{Zchl3} and \eqref{Zchl4} with the proviso above). A final subtlety is that together with $Z_v$ we must transform $\bar V_8$ under $g = RT$ and the modular transformations $S$ and $T$. The first operation acts trivially, and the remaining ones can be read off e.g. from the formulae in Appendix A of \cite{Nakajima:2023zsh}; we obtain
\begin{equation}
	\bar V_8 \overset{g}{\longrightarrow} \bar V_8 \overset{S}{\longrightarrow} \frac12 (\bar O_8 + \bar V_8 - \bar S_8 - \bar C_8) \overset{T}{\longrightarrow} \frac12e^{\pi i /3}(\bar O_8 - \bar V_8 + \bar S_8 + \bar C_8)\,.
\end{equation}
We see in particular that the terms resulting from the operations $g, S, T$ on the vector conjugacy class ultimately contribute to all of the four classes.

Moving on to $Z_s$, we see from \eqref{Zclasses} that the underlying set is completely asymmetric under $R$. This is to be expected since $Z_v+Z_s$ is the partition function (up to Dedekind etas) of the Narain lattice and the corresponding invariant lattice is already contained in the $Z_v$ lattice. As such, $Z_s^{1,g} = 0$ and so $Z_s^{g,1} = Z_s^{g,g} = 0$.

The situation for the twisted partition functions $Z_c$ and $Z_o$ is analogous to the previous ones. The union of the underlying sets is equal to the Narain lattice shifted by the vector $\delta\times \delta = (1,0^7)\times (1,0^7)\in (E_8 \oplus E_8)\otimes \mathbb{R}$. Correspondingly, the invariant set (which is not a lattice since there is no zero element) is $\Gamma_{1,1}\oplus E_8^\delta(2)$ where $E_8^\delta$ is the $E_8$ lattice shifted by $\delta$. We can write the partition function of $E_8^\delta$ as $V_{16}+C_{16}$, and from \eqref{Zclasses} we see that, again, the full set coincides with the invariant subset in $Z_c$. 

Now we perform the necessary transformations. Under $g$, the function $Z_c$ transforms to
\begin{equation}\label{Zc1g}
	Z_c^{1,g} = \frac{1}{\bar \eta \eta} \zeta_2 Z_{S^1}^{(2)} \sum_{p \in E_8(2)} e^{\pi i (p+\sqrt2\delta)^2\tau}\,,
\end{equation}
where the $\sqrt2$ multiplying $\delta$ accounts for the scaling $E_8 \to E_8(2)$. Note that this expression is a shifted version of eq. \eqref{Zchl2}. To obtain $Z_c^{g,1}$ we transform $\tau \to -1/\tau$ and do a Poisson resummation. It suffices to see how the sum over $E_8(2)$ transforms as the rest of the expression is just as for the supersymmetric case. We have that
\begin{equation}
	\sum_{p \in E_8(2)} e^{\pi i (p+\sqrt2\delta)^2\tau} ~~~~~ \overset{S}{\longrightarrow} ~~~~~ \sum_{p \in E_8(2)} e^{-\pi i (p+\sqrt2\delta)^2/\tau}\,.
\end{equation}
Using then the Poisson resummation formula
\begin{equation}
	\sum_{W \in \Lambda}e^{-\pi a (W+U)^2} = \frac{1}{\text{Vol}(\Lambda) a^{\mathcal{D}}}\sum_{P \in \Lambda^*}e^{-\tfrac{\pi}{a}P^2} e^{-2i \pi P \cdot U}
\end{equation}
with $W = p$, $\Lambda = E_8(2)$, $a = i/\tau$, $U = \sqrt2\delta$ and $\mathcal{D} = 8$, we get
\begin{equation}
	\sum_{p \in E_8(2)} e^{-\pi i (p+\sqrt2\delta)^2/\tau} = \frac{1}{16 \tau^{-4}} \sum_{p \in E_8(\tfrac12)}e^{\pi i \tau p^2}e^{-2i\pi p \cdot (\sqrt2\delta)}\,.
\end{equation}
The upshot is that the overall effect of shifting the invariant lattice by $\sqrt 2 \delta $ gets mapped to a phase in the transformed partition function which acts as $(O_{16},S_{16}) \to (O_{16},-S_{16})$. We then have
\begin{equation}\label{afterpoisson}
	Z_c^{g,1} = \frac{1}{\bar\eta \eta}\zeta_3 Z_{S^1}^{(3)}\left(\sum_{p\in D_8(1/2)}q^{\tfrac12 p^2}-\sum_{p\in (E_8/D_8)(1/2)}q^{\tfrac12 p^2}\right)\,,
\end{equation}
cf. eq. \eqref{Zchl3}. 

Finally, $Z_c^{g,g}$ takes the form
\begin{equation}
	Z_c^{g,g} = \frac{1}{\bar\eta \eta}\zeta_4 Z_{S^1}^{(4)}\left(\sum_{p\in D_8(1/2)}q^{\tfrac12 p^2}(-1)^{p^2}-\sum_{p\in (E_8/D_8)(1/2)}q^{\tfrac12 p^2}(-1)^{p^2}\right)\,.
\end{equation}
Again, to see how these functions contribute to the different spacetime classes we must account for the transformations of the character $\bar C_8$,
\begin{equation}
	\bar C_8 \overset{g}{\longrightarrow} \bar C_8 \overset{S}{\longrightarrow} \frac12 (\bar O_8 - \bar V_8 - \bar S_8 + \bar C_8) \overset{T}{\longrightarrow} \frac12e^{\pi i /3}(\bar O_8 + \bar V_8 + \bar S_8 - \bar C_8)\,.
\end{equation}
As in the case of $Z_s$ explained above, we have that $Z_o^{1,g} = Z_o^{g,1} = Z_o^{g,g} = 0$, and so we have all the information needed to write down the full partition function of the orbifold. 

\subsubsection{The partition function}
The vector class receives contributions from $Z_v^{1,1}, Z_v^{1,g}, Z_v^{g,1}, Z_v^{g,g}, Z_c^{g,1}$ and $Z_{g,g}$. Taking into account the transformations of $\bar V_8$ and $\bar C_8$ as well as the relative signs in eq. \eqref{appZ}, we have
\begin{equation}\label{z2chlv}
	Z_v' = \frac{1}{2}\left( Z_v^{1,1} + Z_v^{1,g} + Z_v^{g,1} - Z_v^{g,g} + Z_c^{g,1} - Z_c^{g,g} \right)\,.
\end{equation}
In particular, we have that
\begin{equation}\label{sumcancel1}
	\frac12 (Z_v^{g,1} + Z_c^{g,1}) = \frac{1}{\bar\eta \eta}\zeta_3 Z_{S^1}^{(3)}\sum_{p \in D_8(1/2)}q^{\tfrac12 p^2}\,;
\end{equation}
we see that the $S_{16}$ contributions cancel and we are left with $\eta^8(\tau/2)O_{16} = \sum_{p \in D_8(2)}q^{\tfrac12 p^2}$. Similarly,
\begin{equation}\label{sumcancel2}
	\frac12 (Z_v^{g,g} + Z_c^{g,g}) = \zeta_4 Z_{S^1}^{(4)} \sum_{p \in D_8(1/2)}q^{p^2/2} (-1)^{p^2}\,,
\end{equation}
a result which can be anticipated by noting that this is nothing more than the $T$-transformation of eq. \eqref{sumcancel1}. Therefore the full vector class takes the same form as in the supersymmetric CHL string up to an exchange of the lattice $E_8\oplus E_8$ for $[D_8 \oplus D_8]^+$ in $Z^{1,1}_\Lambda$ and $E_8$ for $D_8$ in $Z_\Lambda^{g,1}, Z_\Lambda^{g,g}$ in eqs. \eqref{Zchl1} to \eqref{Zchl4}.

The spinor class is given by
\begin{equation}\label{z2chls}
	Z_s' = \frac{1}{2}\left( Z_s^{1,1} + Z_v^{g,1} - Z_v^{g,g} - Z_c^{g,1} + Z_c^{g,g} \right)\,,
\end{equation}
which now involves subtractions $Z_{v,c}^{g,1}-Z_{v,c}^{g,g}$, and the contributions read as in eqs. \eqref{sumcancel1} and \eqref{sumcancel2} with the sum over $D_8(1/2)$ replaced by a sum over its complement in $E_8(1/2)$, which we denote $(E_8/D_8)(1/2)$, while $Z_s^{1,1}$ is read from the parent theory data. On the other hand, for the cospinor class we have 
\begin{equation}\label{z2chlc}
	Z_c' = \frac{1}{2}\left( Z_c^{1,1} + Z_c^{1,g} + Z_c^{g,1} - Z_c^{g,g} + Z_v^{g,1} - Z_v^{g,g} \right)\,,
\end{equation}
which as for the vector class involve the sums \eqref{sumcancel1} and \eqref{sumcancel2}, while the first two terms can be read off from the parent theory data and eq. \eqref{Zc1g}. Finally, the scalar class reads
\begin{equation}\label{z2chlo}
	Z_o' = \frac{1}{2}\left( Z_o^{1,1} + Z_v^{g,1} + Z_v^{g,g} - Z_c^{g,1} - Z_c^{g,g} \right)\,.
\end{equation}

To summarize our results, it is convenient to define the functions
\begin{align}
	\mathcal{G}^{(1)}_\Lambda(\tau,\bar\tau) &=  ~~\frac{1}{\bar \eta \eta} Z_{S^1}^{(1)} \sum_{p \in \Lambda}q^{\frac12 p^2} \,,\\
	\mathcal{G}^{(2)}_\Lambda(\tau,\bar \tau) &= ~~\frac{1}{\bar\eta \eta}\zeta_2 Z_{S^1}^{(2)} \sum_{p \in \Lambda(2)}q^{\tfrac12 p^2}\,,\\
	\mathcal{G}^{(3)}_\Lambda(\tau,\bar \tau) &= ~~\frac{1}{\bar\eta \eta}\zeta_3 Z_{S^1}^{(3)}\sum_{p \in \Lambda(1/2)}q^{\tfrac12 p^2}\,,\\
	\mathcal{G}^{(4)}_\Lambda(\tau, \bar \tau) &= \frac{1}{\bar\eta \eta}\zeta_4 Z_{S^1}^{(4)} \sum_{p \in \Lambda (1/2)}q^{p^2/2} (-1)^{p^2}\,.
\end{align}

The partition function of our orbifold then reads
\begin{equation}\label{o16z20}
	Z'(\tau,\bar \tau) = \frac{1}{\tau_2^{7/2} \bar \eta^7 \eta^7}  \times \frac{1}{\eta^{16}}\left(Z_v' \bar V_8 - Z_s' \bar S_8 - Z_c' \bar C_8 + Z_o' \bar O_8   \right)\,,
\end{equation}
with
\begin{align}
	Z_v' &= \mathcal{G}^{(1)}_{\Gamma_v} + \mathcal{G}^{(2)}_{E_8} + \mathcal{G}^{(3)}_{D_8} + \mathcal{G}^{(4)}_{D_8}\,,\\
	Z_s' &= \mathcal{G}^{(1)}_{\Gamma_s}  + \mathcal{G}^{(3)}_{E_8/D_8} + \mathcal{G}^{(4)}_{E_8/D_8}\,,\\
	Z_c' &= \mathcal{G}^{(1)}_{\Gamma_v+\delta} + \mathcal{G}^{(2)}_{E_8+\delta} + \mathcal{G}^{(3)}_{D_8} + \mathcal{G}^{(4)}_{D_8}\,,\\
	Z_o' &= \mathcal{G}^{(1)}_{\Gamma_s+ \delta}  + \mathcal{G}^{(3)}_{E_8/D_8} - \mathcal{G}^{(4)}_{E_8/D_8}\,,\label{o16z24}
\end{align}

where $\Gamma_v \equiv [2\, D_8]^+$, $\Gamma_s \equiv 2\,E_8/\Gamma_v$, and $\Gamma_{v,s}+\delta$ are shifted sets with $\delta = (1,0^7)$.

\subsubsection{T-duality}\label{sss:o16t}
Now we wish to show that upon turning on a Wilson line in the non-supersymmetric CHL string we can interpolate with this theory. To this end let us set $R = R_0 = \sqrt{2}$ in our theory so that the gauge group is fully enhanced. The momenta read
\begin{equation}
	p_{L,R} = \frac{1}{\sqrt2}\left(\frac{n}{R_0} \pm wR_0  \right) = \frac{n}{2} \pm w\,.
\end{equation}
The number of level matched states for the first few levels are then given by the expansions
\begin{equation}\label{expTd}
	\begin{split}
		\frac{1}{\bar \eta^7 \eta^{23}}Z_v' \bar V_8 &= 1040 + 3840 (\bar q q)^{1/8} + 37440 (\bar q q)^{1/2} + \cdots \,, \\
		\frac{1}{\bar \eta^7 \eta^{23}}Z_s' \bar S_8 &= 1024 + 4096 (\bar q q)^{1/8} + 36864 (\bar q q)^{1/2} + \cdots \,, \\
		\frac{1}{\bar \eta^7 \eta^{23}}Z_c' \bar C_8 &= 1104 + 3840 (\bar q q)^{1/8} + 37696 (\bar q q)^{1/2} + \cdots \,, \\
		\frac{1}{\bar \eta^7 \eta^{23}}Z_o' \bar O_8 &= 256 + 82944 (\bar q q)^{1/2} + \cdots \,. \\
	\end{split}
\end{equation}
The 1040 massless states in the first line include $123\times 8 = 984$ corresponding to the gauge algebra $\mathfrak{g} = \mathfrak{so}_{16}\oplus \mathfrak{su}_2$, the rest furnishing the usual ``gravity multiplet" fields. There is also a spinor transforming in the $\mathbf{128}$ of $\mathfrak{so}_{16}$, a cospinor in the $\mathbf{135}$ of $\mathfrak{so}_{16}$ (traceless symmetric) and a cospinor in the adjoint of $\mathfrak{su}(2)$. Finally there are 256 massless scalars signaling an instability of the ``knife edge type" \cite{Ginsparg:1986wr,Fraiman:2023cpa}.

The expansions \eqref{expTd} can be reproduced in the theory considered in the previous subsection by a suitable choice of Wilson line and radius moduli furnishing the gauge algebra $\mathfrak{so}_{16}\oplus \mathfrak{su}_2$, although the computations are somewhat intensive.\footnote{We thank B. Fraiman for assistance on this point.} This is a clear indication that both theories are T-dual, and furthermore serves as a consistency check for the computation of the partition function just carried out. The details of how the states arrange into various representations of the gauge algebra can be deduced using the results of \cite{Nakajima:2023zsh}, and we have carried over that information to this case. 

We can moreover compactify the theory down to two dimensions and factorize the partition function. This is easily done along the same lines as for the CHL string, and a $q$-expansion yields 
\begin{equation}
	\vec Z_L^t = 
	\begin{pmatrix}
		q^{-1} + 256 + 98580 q + O(q^2)\\
		128 + 98304q + O(q^2)\\
		152 + 98304q + O(q^2)\\
		4096q^{1/2} + O(q^{3/2})
	\end{pmatrix}\,,
\end{equation}
matching perfectly with the data of item 519 in Table 12 of \cite{Hohn:2023auw}, which sits inside genus $\mathcal{B}_{IIb}$ as advertised.

\subsection{Scherk-Schwarz reduction of the CHL string}\label{ss:SSCHL}
Let us now consider the Scherk-Schwarz reduction of the CHL string, which following the discussion in Section \ref{ss:nonsusychl}, should match with the genus $\mathcal{B}_I$ of \cite{Hohn:2023auw}. Computing the partition function turns out to be easier by considering instead the Chauduri-Polchinski orbifold of the Scherk-Schwarz reduction of the $E_8 \times E_8$ string compactified on another circle. Both procedures are equivalent, as the orbifolds commute, but the latter allows us to proceed in a manner analogous to that of the previous subsection.

The orbifolding procedure is straightforward, but let us record here how it is done schematically. The partition function of the Scherk-Schwarz reduction of the $E_8 \times E_8$ string, obtained with $g =  T(-1)^F$, has four contributions (with $Z_i \equiv Z_{S^1}^{(i)}$)
\begin{equation}
	\begin{split}
		Z^{1,1} &\sim  Z_1 \times (O_{16}+S_{16})^2 (\bar V_8 - \bar S_8)\,,\\
		Z^{1,g} &\sim  Z_2 \times (O_{16}+S_{16})^2 (\bar V_8 + \bar S_8)\,,\\
		Z^{g,1} &\sim  Z_3 \times (O_{16}+S_{16})^2 (\bar O_8 - \bar C_8)\,,\\
		Z^{g,g} &\sim  Z_4 \times (O_{16}+S_{16})^2 (\bar O_8 + \bar C_8)\,.
	\end{split}
\end{equation}
Compactifying further on a circle adds a $Z_1$ factor to every term, and starting from each one of them we can then perform the Chaudhuri-Polchinski orbifold with $h = RT'$ with $T'$ the half-shift along the new circle. Starting from $Z^{1,1}$ we obtain three extra terms yielding in total the four contributions to the supersymmetric CHL string in eight dimensions,
\begin{equation}
	\begin{split}
		Z^{1,1}_{1,1} &\sim  Z_1 \times Z_1 \times  (O_{16}+S_{16})^2 (\bar V_8 - \bar S_8)\,,\\
		Z^{1,1}_{1,h} &\sim  Z_2 \times Z_1 \times \zeta_2 (O_{16}(2\tau)+S_{16}(2\tau)) (\bar V_8 - \bar S_8)\,,\\
		Z^{1,1}_{h,1} &\sim  Z_3 \times Z_1 \times \zeta_3 (O_{16}(\tau/2)+S_{16}(\tau/2)) (\bar V_8 - \bar S_8)\,,\\
		Z^{1,1}_{h,h} &\sim  Z_4 \times Z_1 \times \zeta_4 (O_{16}'(\tau/2)+S_{16}'(\tau/2))(\bar V_8 - \bar S_8)\,.
	\end{split}
\end{equation}
For the remaining terms we need to consider the transformation properties of the corresponding spacetime character combinations as well as the factors $Z_i$. We obtain 
\begin{equation}
	\begin{split}
		Z^{1,g}_{1,1} &\sim  Z_1 \times Z_2 \times  (O_{16}+S_{16})^2 (\bar V_8 + \bar S_8)\,,\\
		Z^{1,g}_{1,h} &\sim  Z_2 \times Z_2 \times \zeta_2 (O_{16}(2\tau)+S_{16}(2\tau)) (\bar V_8 + \bar S_8)\,,\\
		Z^{1,g}_{h,1} &\sim  Z_3 \times Z_3 \times \zeta_3 (O_{16}(\tau/2)+S_{16}(\tau/2)) (\bar O_8 - \bar C_8)\,,\\
		Z^{1,g}_{h,h} &\sim  Z_4 \times Z_4 \times \zeta_4 (O_{16}'(\tau/2)+S_{16}'(\tau/2)) (\bar O_8 + \bar C_8)\,,
	\end{split}
\end{equation}
\begin{equation}
	\begin{split}
		Z^{g,1}_{1,1} &\sim  Z_1 \times Z_3 \times (O_{16}+S_{16})^2 (\bar O_8 - \bar C_8)\,,\\
		Z^{g,1}_{1,h} &\sim  Z_2 \times Z_3 \times \zeta_2 (O_{16}(2\tau)+S_{16}(2\tau)) (\bar O_8 - \bar C_8)\,,\\
		Z^{g,1}_{h,1} &\sim  Z_3 \times Z_2 \times \zeta_3 (O_{16}(\tau/2)+S_{16}(\tau/2)) (\bar V_8 + \bar S_8)\,,\\
		Z^{g,1}_{h,h} &\sim  Z_4 \times Z_2 \times \zeta_4 (O_{16}'(\tau/2)+S_{16}'(\tau/2)) (\bar V_8 + \bar S_8)\,,
	\end{split}
\end{equation}
\begin{equation}
	\begin{split}
		Z^{g,g}_{1,1} &\sim  Z_1 \times Z_4 \times (O_{16}+S_{16})^2 (\bar O_8 + \bar C_8)\,,\\
		Z^{g,g}_{1,h} &\sim  Z_2 \times Z_4 \times \zeta_2 (O_{16}(2\tau)+S_{16}(2\tau)) (\bar O_8 + \bar C_8)\,,\\
		Z^{g,g}_{h,1} &\sim  Z_3 \times Z_4 \times \zeta_3 (O_{16}(\tau/2)+S_{16}(\tau/2)) (\bar O_8 + \bar C_8)\,,\\
		Z^{g,g}_{h,h} &\sim  Z_4 \times Z_3 \times \zeta_4 (O_{16}'(\tau/2)+S_{16}'(\tau/2)) (\bar O_8 - \bar C_8)\,.
	\end{split}
\end{equation}
We can then put together the contributions to each spacetime class, which involves combinations of the type $Z_1 \pm Z_2$ and $Z_3 \pm Z_4$, amounting to projections of the underlying $\Gamma_{1,1}$ momentum lattice and its shifted version with half-integer winding numbers.

Compactifying on an extra $T^6$ down to two dimensions and factorizing is less straightforward than in the previous cases where only one shift vector was present. The reason is that now we have to account for how the effect of the two shifts overlap on the $E_8$ lattice, and the problem is purely technical. Let us illustrate this with the spacetime vector class, which receives the contributions  
\begin{equation}\label{vcontr}
	\begin{split}
		Z_v  &\sim  Z_1 \times (Z_1 + Z_2)\times  (O_{16}+S_{16})^2 \\
		&+  Z_2 \times (Z_1 + Z_2) \times \zeta_2 (O_{16}(2\tau)+S_{16}(2\tau)) \\
		&+  Z_3 \times (Z_1 + Z_2) \times \zeta_3 (O_{16}(\tau/2)+S_{16}(\tau/2)) \\
		&+  Z_4 \times (Z_1 + Z_2) \times \zeta_4 (O_{16}'(\tau/2)+S_{16}'(\tau/2))\,.
	\end{split}
\end{equation}
Compactifying and polarizing the underlying lattice will be equivalent to choosing how the shift vectors and projection will act on the $E_8$ lattice. In this case, the contributions $Z_1,...,Z_4$ give rise to a situation as in the CHL string, where se saw that the analogs for the $E_8$ are given by the character combinations
\begin{equation}
	(O_{16} + S_{16})\,, ~~~~~ 	(O_{16} - S_{16})\,, ~~~~~ 	(V_{16} + C_{16})\,, ~~~~~ 	(V_{16} - C_{16})\,,
\end{equation}
respectively. The corresponding shift is $\delta' = (1,0^7) \in E_8\otimes \mathbb{R}$, which selects out a $D_8$ sublattice. We take the shift associated to the Scherk-Schwarz reduction circle to be $\delta = (\tfrac12^4,0^4)$ which selects an index 2 sublattice of $D_8$, namely $2\, D_4$. We need then to project out the contributions to the characters above which come from vectors not in $2\, D_4 \in D_8$ and its associated weights corresponding to the different conjugacy classes.  The projection maps 
\begin{equation}
	(O_{16}, S_{16}, C_{16}, V_{16}) \to (O_{8}^2, S_8^2, C_8S_8,O_8V_8)\,,
\end{equation}
and the terms in \eqref{vcontr} transform to
\begin{equation}
	\begin{split}
		Z_v  &= \frac14  (O_8^2 + S_8^2) (O_{16}+S_{16})^2 \\
		&+  \frac14\zeta_2(O_8^2 - S_8^2)  (O_{16}(2\tau)+S_{16}(2\tau)) \\
		&+  \frac14\zeta_3(C_8 S_8 + V_8 O_8 )  (O_{16}(\tau/2)+S_{16}(\tau/2)) \\
		&+  \frac14\zeta_4(C_8 S_8 - V_8 O_8) (O_{16}'(\tau/2)+S_{16}'(\tau/2))\,,
	\end{split}
\end{equation}
where all factors are now accounted for, and in particular the $1/4$ prefactors come from the relation $Z = \frac12 (Z^{1,1}+\cdots+Z^{g,g})$ for $g,h$. The procedure is analogous for the remaining classes, and in the end we obtain a left-moving CFT whose partition function is $q$-expanded as 
\begin{equation}
	\vec Z_L^t = 
	\begin{pmatrix}
		q^{-1} + 312 + 114964 q + O(q^2)\\
		72 + 10354688q + O(q^2)\\
		72 + 10354688q + O(q^2)\\
		8q^{-1} + 6304q^{1/2} + O(q^{3/2})
	\end{pmatrix}\,,
\end{equation}
corresponding exactly to item 274 of \cite{Hohn:2023auw}. As expected, this item lies in genus $\mathcal{B}_I$. 

\subsection{Scherk-Schwarz reduction of the $E_8$ string}
Now we seek to construct a theory which matches with the genus $\mathcal{B}_{IIa}$. One interesting avenue to try is performing a Scherk-Schwarz reduction of another non-supersymmetric theory. Usually we would employ this method to break supersymmetry, but more generally it can also produce one non-supersymmetric theory from another. We will find that starting from the $E_8$ string, the theory obtained through a Scherk-Schwarz reduction is a new one which turns out to be in correspondence with genus $\mathcal{B}_{IIa}$, thus completing the classification of non-supersymmetric CHL-like theories. In the next subsection we will see that another natural construction given by the Chaudhuri-Polchinski orbifold of the $E_7 \times SU(2) \times E_7 \times SU(2)$ string is also in correspondence to genus $\mathcal{B}_{IIa}$ and is thus T-dual to the former. 

As before, it is simpler to start from the Scherk-Schwarz reduction of the $E_8 \times E_8$ string and then perform the orbifold defining the original theory, this time given by $R(-1)^F$. The story is similar to that of the previous section, but now we stay in nine dimensions. Proceeding schematically, we start again with 
\begin{equation}
	\begin{split}
		Z^{1,1} &\sim  Z_1 \times (O_{16}+S_{16})^2 (\bar V_8 - \bar S_8)\,,\\
		Z^{1,g} &\sim  Z_2 \times (O_{16}+S_{16})^2 (\bar V_8 + \bar S_8)\,,\\
		Z^{g,1} &\sim  Z_3 \times (O_{16}+S_{16})^2 (\bar O_8 - \bar C_8)\,,\\
		Z^{g,g} &\sim  Z_4 \times (O_{16}+S_{16})^2 (\bar O_8 + \bar C_8)\,.
	\end{split}
\end{equation}
Applying the orbifold with $h = R (-1)^F$ we obtain from $Z^{1,1}$ the circle compactification of the $E_8$ string (cf. eq. \eqref{ZE8}),
\begin{equation}
	\begin{split}
		Z^{1,1}_{1,1} &\sim Z_1 \times  (O_{16}+S_{16})^2 (\bar V_8 - \bar S_8)\,,\\
		Z^{1,1}_{1,h} &\sim Z_1 \times \zeta_2 (O_{16}(2\tau)+S_{16}(2\tau)) (\bar V_8 + \bar S_8)\,,\\
		Z^{1,1}_{h,1} &\sim Z_1 \times \zeta_3 (O_{16}(\tau/2)+S_{16}(\tau/2)) (\bar O_8 - \bar C_8)\,,\\
		Z^{1,1}_{h,h} &\sim Z_1 \times \zeta_4 (O_{16}'(\tau/2)+S_{16}'(\tau/2))(\bar O_8 + \bar C_8)\,.
	\end{split}
\end{equation}
The next three terms give rise to
\begin{equation}
	\begin{split}
		Z^{1,g}_{1,1} &\sim Z_2 \times  (O_{16}+S_{16})^2 (\bar V_8 + \bar S_8)\,,\\
		Z^{1,g}_{1,h} &\sim Z_2 \times \zeta_2 (O_{16}(2\tau)+S_{16}(2\tau)) (\bar V_8 - \bar S_8)\,,\\
		Z^{1,g}_{h,1} &\sim Z_3 \times \zeta_3 (O_{16}(\tau/2)+S_{16}(\tau/2)) (\bar V_8 - \bar S_8)\,,\\
		Z^{1,g}_{h,h} &\sim Z_4 \times \zeta_4 (O_{16}'(\tau/2)+S_{16}'(\tau/2))(\bar V_8 - \bar S_8)\,,
	\end{split}
\end{equation}
\begin{equation}
	\begin{split}
		Z^{g,1}_{1,1} &\sim Z_3 \times  (O_{16}+S_{16})^2 (\bar O_8 - \bar C_8)\,,\\
		Z^{g,1}_{1,h} &\sim Z_3 \times \zeta_2 (O_{16}(2\tau)+S_{16}(2\tau)) (\bar O_8 + \bar C_8)\,,\\
		Z^{g,1}_{h,1} &\sim Z_2 \times \zeta_3 (O_{16}(\tau/2)+S_{16}(\tau/2))  (\bar O_8 + \bar C_8)\,,\\
		Z^{g,1}_{h,h} &\sim Z_2 \times \zeta_4 (O_{16}'(\tau/2)+S_{16}'(\tau/2)) (\bar O_8 - \bar C_8)\,,
	\end{split}
\end{equation}
\begin{equation}
	\begin{split}
		Z^{g,g}_{1,1} &\sim Z_4 \times  (O_{16}+S_{16})^2 (\bar O_8 + \bar C_8)\,,\\
		Z^{g,g}_{1,h} &\sim Z_4 \times \zeta_2 (O_{16}(2\tau)+S_{16}(2\tau)) (\bar O_8 - \bar C_8)\,,\\
		Z^{g,g}_{h,1} &\sim Z_4 \times \zeta_3 (O_{16}(\tau/2)+S_{16}(\tau/2))  (\bar V_8 + \bar S_8)\,,\\
		Z^{g,g}_{h,h} &\sim Z_3 \times \zeta_4 (O_{16}'(\tau/2)+S_{16}'(\tau/2)) (\bar V_8 + \bar S_8)\,.
	\end{split}
\end{equation}

Putting together the contributions to the different spacetime conjugacy classes, we get again projections of the form $Z_1 \pm Z_2$ and $Z_3 \pm Z_4$, but since there is no extra shift, compactification and factorization is more straightforward than in the case of section \ref{ss:SSCHL}. We choose a shift splitting $E_8$ into $D_8$ and its spinor class as for the CHL string, such that we have the correspondences
\begin{equation}
	Z_1 + Z_2 \leftrightarrow O_{16}\,, ~~~~~ Z_1 - Z_2 \leftrightarrow S_{16}\,, ~~~~~ Z_3 + Z_4 \leftrightarrow V_{16}\,, ~~~~~ Z_3 - Z_4 \leftrightarrow  C_{16}\,.
\end{equation}
Again as an example we have the vector class taking the form
\begin{equation}
	\begin{split}
		Z_v  &= \frac14  O_{16}(O_{16}+S_{16})^2 \\
		&+  \frac14\zeta_2O_{16}  (O_{16}(2\tau)+S_{16}(2\tau)) \\
		&+  \frac14\zeta_3V_{16}(O_{16}(\tau/2)+S_{16}(\tau/2)) \\
		&-  \frac14\zeta_4V_{16} (O_{16}'(\tau/2)+S_{16}'(\tau/2))\,.
	\end{split}
\end{equation}
Putting everything together we end up with a left-moving CFT whose partition function has $q$-expansion
\begin{equation}
	\vec Z_L^t = 
	\begin{pmatrix}
		q^{-1} + 384 + 133396 q + O(q^2)\\
		63488q + O(q^2)\\
		63488q + O(q^2)\\
		17q^{-1} + 8788q^{1/2} + O(q^{3/2})
	\end{pmatrix}\,,
\end{equation}
which matches item 448 of \cite{Hohn:2023auw}, the first one in genus $\mathcal{B}_{IIa}$. 
\subsection{$\mathbb{Z}_2$ orbifold of the $E_7\times SU(2)\times E_7\times SU(2)$ string}\label{ss:CPE7}
Finally we wish to consider another ten dimensional non-supersymmetric theory admitting a Chaudhuri-Polchinski orbifold upon circle compactification, namely the $E_7 \times SU(2) \times E_7 \times SU(2)$ heterotic string. We have already completed the classification of theories corresponding to fermionic CFT genera and so we expect that this construction is T-dual to one of those already studied. Indeed, as we will see, we will find a correspondence with genus $\mathcal{B}_{IIa}$, concluding that the theory is T-dual to the Scherk-Schwarz reduction of the $E_8$ string. 

The starting point is the partition function of the $E_7 \times SU(2) \times E_7 \times SU(2)$ compactified on $S^1$, which reads just as for the $O(16)\times O(16)$ string case (cf. eq. \eqref{appZ}) differing in the form of $Z_{v,s,c,o}$. In general, 
\begin{equation}
	\begin{split}
		Z_{v,s,c,o} = \frac{1}{\bar \eta \eta}Z_{S^1}^{(1)}\frac{1}{\eta^{16}}\sum_{p \in \Gamma_{v,s,c,o}}q^{\tfrac12 p^2}\,,
	\end{split}
\end{equation}
where $\Gamma_{v,s,c,o}$ are defined by the shift $\delta$ producing the ten dimensional theory from the $E_8 \times E_8$ string \cite{Ginsparg:1986wr}. In the case of the $O(16)\times O(16)$ string, the resulting $Z_{v,s,c,o}$ can be written simply in terms of $SO(16)$ characters as in eq. \eqref{o16z}. In the case at hand we find it convenient to work in term of the classes $\Gamma_{v,s,c,o}$. The shift vector acting on the lattice $E_8\oplus E_8$ is $\delta = (0^6,\tfrac12,\tfrac12)\times(0^6,\tfrac12,\tfrac12)$, and the different classes are defined as
\begin{equation}
	\begin{split}
		\Gamma_v &= \{ u \in E_8 \oplus E_8 | u\cdot \delta \in \mathbb{Z} \}\,,\\
		\Gamma_s &= \{ u \in E_8 \oplus E_8 | u\cdot \delta \in \mathbb{Z} + 1/2\}\,,\\
		\Gamma_c &= \Gamma_s + \delta\,,\\
		\Gamma_o &= \Gamma_v + \delta\,.
	\end{split}
\end{equation}
Note that for the $O(16)\times O(16)$, where $\delta = (1,0^7)\times (1,0^7)$ has even norm, the last two classes are defined as $\Gamma_c = \Gamma_v + \delta$ and $\Gamma_o = \Gamma_s + \delta$. For $\delta$ with odd norm, the corresponding definitions are as above, and indeed the $E_7\times SU(2) \times E_7 \times SU(2)$ string has tachyons associated to vectors with odd norm in the scalar class $\Gamma_0$ such as $\delta \in \Gamma_0$ itself. 

The orbifolding procedure is largely the same as for the case of the $O(16)\times O(16)$ string carried our in section \ref{ss:CPO16}. On one hand, the lattice invariant under the exchange of gauge factors $R$ is again $\Gamma_{1,1}\oplus E_8(2)$. On the other, the non-trivial step of doing Poisson resummation on the shifted lattice also results in an overall change of sign of the contribution affected by $\delta$ (cf. eq. \eqref{afterpoisson}). There is however the important subtlety that the co-spinor and scalar classes $\Gamma_c$ and $\Gamma_o$ are formally exchanged, and so the overall contributions to the spacetime classes in the final theory read somewhat differently. In this case we must use the transformations
\begin{equation}
	\bar O_8 \overset{g}{\longrightarrow} \bar O_8 \overset{S}{\longrightarrow} \frac12 (\bar O_8 + \bar V_8 + \bar S_8 + \bar C_8) \overset{T}{\longrightarrow} \frac12e^{\pi i /3}(\bar O_8 - \bar V_8 - \bar S_8 - \bar C_8)\,,
\end{equation}
and the overall contributions take the form (cf. eqs. \eqref{z2chlv}, \eqref{z2chls}, \eqref{z2chlc}, \eqref{z2chlo})
\begin{equation}
	\begin{split}
		Z_v' &= \frac{1}{2}\left( Z_v^{1,1} + Z_v^{1,g} + Z_v^{g,1} - Z_v^{g,g} + Z_o^{g,1} - Z_o^{g,g}\right)\,,\\
		Z_s' &= \frac{1}{2}\left( Z_s^{1,1} + Z_v^{g,1} - Z_v^{g,g} - Z_o^{g,1} + Z_o^{g,g} \right)\,,\\
		Z_c' &= \frac{1}{2}\left( Z_c^{1,1}  + Z_v^{g,1} - Z_v^{g,g} - Z_o^{g,1} + Z_o^{g,g} \right)\,,\\
		Z_o' &= \frac{1}{2}\left( Z_o^{1,1} + Z_o^{1,g} + Z_o^{g,1} + Z_o^{g,g} + Z_v^{g,1} + Z_v^{g,g} \right)\,.
	\end{split}
\end{equation}
Note the similarity between $Z_c'$ and $Z_s'$. We know that for the tachyonic heterotic theories the spinor and cospinor classes are equivalent \cite{BoyleSmith:2023xkd,Fraiman:2023cpa} and so $Z_c = Z_s$ hence $Z_c' = Z_s'$, suggesting that our present orbifold is also tachyonic. Indeed, the Chaudhuri-Polchinski orbifold preserves two of the four original tachyons in the theory. We will say more about this below.

In terms of the functions $\mathcal{G}^{(i)}$, the different spacetime classes read
\begin{align}
	Z_v' &= \mathcal{G}^{(1)}_{\Gamma_v} + \mathcal{G}^{(2)}_{E_8} + \mathcal{G}^{(3)}_{(E_7+A_1)} + \mathcal{G}^{(4)}_{(E_7+A_1)}\,,\\
	Z_s' = Z_c' &= \mathcal{G}^{(1)}_{\Gamma_s}  + \mathcal{G}^{(3)}_{E_8/(E_7+A_1)} + \mathcal{G}^{(4)}_{E_8/(E_7+A_1)}\,,\\
	Z_o' &= \mathcal{G}^{(1)}_{\Gamma_v+ \delta} + \mathcal{G}^{(2)}_{E_8+\delta} + \mathcal{G}^{(3)}_{(E_7+A_1)} - \mathcal{G}^{(4)}_{(E_7+A_1)}\,.
\end{align}
Going to the self-dual radius $R = R_0 = \sqrt{2}$, we obtain the (level-matched part of) the $(\bar q,q)$-expansions
\begin{equation}
	\begin{split}
		\frac{1}{\bar \eta^7 \eta^{23}}Z_v' \bar V_8 &= 1168 + 3328 (\bar q q)^{1/8} + 22720 (\bar q q)^{1/2} + \cdots \,, \\
		\frac{1}{\bar \eta^7 \eta^{23}}Z_s' \bar S_8 &= 896 + 4608 (\bar q q)^{1/8} + 53632 (\bar q q)^{1/2} + \cdots \,, \\
		\frac{1}{\bar \eta^7 \eta^{23}}Z_c' \bar C_8 &= 896 + 4608 (\bar q q)^{1/8} + 53632 (\bar q q)^{1/2} + \cdots \,, \\
		\frac{1}{\bar \eta^7 \eta^{23}}Z_o' \bar O_8 &= 3(\bar q q)^{-1/2} + 116 + 73008 (\bar q q)^{1/2} + \cdots \,. \\
	\end{split}
\end{equation}
At this enhancement point, the gauge algebra reads $\mathfrak{e}_7 \oplus \mathfrak{su}_2 \oplus \mathfrak{su}_2$, there is a spinor as well as a co-spinor transforming in the representation $(\mathbf{56},\mathbf{2})$ of $\mathfrak{e}_7 \oplus \mathfrak{su}_2$, and there are three tachyons. Two of these tachyons correspond to the diagonal combinations of the four tachyons in the parent theory, while the third appears in the twisted sector and has moduli-independent mass (i.e. it appears in the whole classical moduli space). The 116 spacetime scalars could signal a further knife-edge instability, but we have not checked this; in any case, the theory is not free of tree-level tachyons. 

The fact that there is a moduli-independent tachyon is related to the fact that this theory is T-dual to the Scherk-Schwarz reduction of the $E_8$ string. The $E_8$ string tachyon cannot be removed by compactifying on a circle and turning on Wilson lines, hence it is moduli-independent, and the Scherk-Schwarz reduction has no effect on this tachyon. To see why these theories are T-dual, we compactify the orbifold just constructed down to two dimensions and polarize the lattice as done previously. We find the left-moving CFT
\begin{equation}
	\vec Z_L^t = 
	\begin{pmatrix}
		q^{-1} + 272 + 104724 q + O(q^2)\\
		112 + 92160q + O(q^2)\\
		112 + 92160q + O(q^2)\\
		3q^{-1/2} + 4924q^{1/2} + O(q^{3/2})
	\end{pmatrix}\,,
\end{equation}
matching item 449 of Table 12 in \cite{Hohn:2023auw}, which lies in genus $\mathcal{B}_{IIa}$. Both theories can be marginally deformed in 2D to worldsheets with left-moving CFTs in genus $\mathcal{B}_{IIa}$ and since we have argued that all such theories are connected classically, T-duality follows. We could attempt to explicitly match both theories by turning on Wilson lines in one of them (as we have done previously in section \ref{sss:o16t}), but as already commented this computation is intensive and we do not find it worthwhile.

\subsection{Comments on tree-level tachyons}
Even though we have limited our attention to an explicit study of theories with rank reduced by 8, it is reasonable to expect that this case captures the most salient features of non-supersymmetric strings with rank reduction. In particular, we have seen that there are four types of fermionic CFT genera and all of these can be realized for rank reduced by 8. The next question is then what are the qualitative differences between these four string theories. In this section we will limit ourselves to one such qualitative difference which concerns the way in which tree-level tachyons appear in the spectrum as one moves in the classical moduli space.  

This qualitative distinction can be anticipated at a glance from the data in Table 12 of \cite{Hohn:2023auw}. Each fermionic CFT is classified by three integers, $a$, $b$ and $\ell$, and it turns out that $\ell$ is equal to the number of tachyons with minimal squared mass (extremal tachyons in the following) in the associated heterotic worldsheet. One can then note the curious pattern that $\ell$ takes distinct definite values for each genus:
\begin{equation}
	\ell = \begin{cases}
		\text{even} & \mathcal{B}_{I}\\
		\text{odd} & \mathcal{B}_{IIa}\\
		0 & \mathcal{B}_{IIb}\\
		1 & \mathcal{B}_{III}
	\end{cases}\,.
\end{equation} 
The first case can be explained in analogy with genus $\mathcal{A}_I$. For this theory, tachyons are in correspondence with vectors with norm $1$ in the spacetime scalar conjugacy class. Since vectors always come in pairs $\pm v$, it follows that tachyons will come in pairs. A preliminar analysis of genus $\mathcal{B}_{IIa}$ shows on the other hand that this behavior is also present, but furthermore there is a generic, moduli-independent tachyon, explaining the pattern. For genus $\mathcal{B}_{IIb}$, extremal tachyons correpond to null vectors in the charge lattice, and so they can only appear at infinite distance, where they form an infinite tower of tachyonic states corresponding e.g. to the $E_8$ string tachyon in ten dimensions. Finally, genus $\mathcal{B}_{III}$ behaves in a similar way with the difference that there is a generic, moduli-independent tachyon. One should keep in mind that extra tachyons do appear at finite distance in the classical moduli space in genera $\mathcal{B}_{IIb}$ and $\mathcal{B}_{III}$, as their mass crosses the threshold where they become massless. In particular, many maximal enhancements for genus $\mathcal{B}_{IIb}$ do have tree-level tachyons; they are simply not extremal. 

The discussion so far is reasonably intuitive. Extremal tachyons are either always there or are associated to pairs of vectors in the charge lattice of the theory. The careful reader will note, however, that there is some data in Table 12 of \cite{Hohn:2023auw} going against this intuition. Namely, if we look at genus $\mathcal{D}_{IIa}$, the number of extremal tachyons can be either $0, 1, 2$ or $3$. The orbifold associated to this genus exhibits order doubling \cite{Hohn:2020xfe} and that could account for some novel effect explaining this behavior. Indeed, genera $\mathcal{J}_{IIa}$ and $\mathcal{K}_{IIa}$, which also exhibit order doubling, behave in this way too. More interestingly, genus $\mathcal{D}_{III}$ presents either 0 or 1 extremal tachyons at the holomorphic factorization points, which goes against what we would expect given its formal relationship with genus $\mathcal{B}_{III}$, i.e. the $E_8$ string. We will come back to this problem in the future.

\section{The remaining non-supersymmetric models}\label{s:remaining}
We have seen thus far that the CFT genera $\mathcal{A}_I, \mathcal{B}_{I}, \mathcal{B}_{IIa}, \mathcal{B}_{IIb}$ and $\mathcal{B}_{III}$ are in perfect correspondence with non-supersymmetric heterotic strings living respectively in $10, 8, 9, 9$ and $10$ dimensions at most. There are of course many other genera which could lead to non-supersymmetric strings with lower ranks, and in this section we discuss to what extent we know this correspondence to hold. 

\subsection{Scherk-Schwarz reduction of supersymmetric strings}
As we have seen, genera $\mathcal{A}_I$ and $\mathcal{B}_I$ correspond respectively to the Scherk-Schwarz reductions of the rank 16 supersymmetric heterotic strings and the CHL strings. From Table 7 of \cite{Hohn:2023auw} we see that there are in total 7 genera of this type:
\begin{equation}
	\text{Type I genera:} ~~~~~ \mathcal{A}\,, ~~~ \mathcal{B}\,, ~~~ \mathcal{C}\,, ~~~ \mathcal{E}\,, ~~~ \mathcal{F}\,, ~~~ \mathcal{G}\,, ~~~ \mathcal{H}\,.
\end{equation}
On the other hand we know that these letters correspond to the supersymmetric theories realized respectively by $\mathbb{Z}_1\,,...\,, \mathbb{Z}_6$-heterotic-triples and the $\mathbb{Z}_7$-heterotic quadruple (the first two of course corresponding to the two theories above). The heterotic triples can be realized in seven dimensions, and so by the Scherk-Schwarz mechanism we realize corresponding six dimensional non-supersymmetric theories. As for genus $\mathcal{H}_I$, the corresponding supersymmetric theory is defined in up to six dimensions and its Scherk-Schwarz reduction leads to a five dimensional theory. We do not know if such a theory admits a non-trivial decompactification limit to six dimensions, but it would be interesting to check this. 

A non-trivial observation is that we do know that there is a $\mathbb{Z}_8$-heterotic quadruple in six dimensions, but Hohn and Möller have not found a corresponding $\mathcal{I}_I$ genus. This suggests that, at least for these theories, there is indeed a canonical correspondence between the type I genera and theories living in six or more dimensions, hence the decompactification limit mentioned at the end of the previous paragraph is likely to exist. 

\subsection{Alternative factorization, MSDS models}
Another important aspect of Scherk-Schwarz reductions is that upon compactification to two dimensions, they admit factorization points in their classical moduli spaces which differ from those studied previously. At these points, the right-moving part of the worldsheet is not the sCFT based on the $E_8$ lattice but rather the CFT consisting on 24 real fermions with current algebra $8\, \mathfrak{su}_2$. 

This can be nicely illustrated using genus $\mathcal{A}_I$. As discussed in section \ref{ss:rk16}, the dual of the charge lattice in two dimensions can be polarized to $\Upsilon_{24,8}^* \simeq L_i \oplus E_8(-1)$, where $L_i$ is a lattice corresponding to any of the 273 fermionic CFTs in genus $\mathcal{A}_I$. We can choose, for example,
\begin{equation}\label{e8d8}
	\Upsilon_{24,8}^* \simeq E_8 \oplus E_8 \oplus D_8 \oplus E_8(-1)\,.
\end{equation}
However, it is easy to show that $D_8 \oplus E_8(-1)$ is isomorphic to $E_8 \oplus D_8(-1)$ (both have $\Gamma_{1,1}$ sublattice and share the quadratic discriminant form \cite{NikulinVV1980ISBF}, see alternative explanation below), which corresponds to another factorization point with partition function
\begin{equation}
	Z(\bar \tau, \tau) = (O_{16}+S_{16})^3 (\bar O_{16}\bar V_8 - \bar S_{16}\bar C_8 - \bar C_{16}\bar S_8 + \bar V_{16}\bar O_8)\,.
\end{equation} 
The right-moving contribution can be checked to be numerically equal to 24, and from an analysis of the spectrum of the theory \cite{Fraiman:2023cpa} we know that the underlying lattice $D_8$ corresponds to a spacetime gauge algebra $8\, \mathfrak{su}_{2,2}$ (more precisely the gauge group is $SU(2)^8/\mathbb{Z}_2^7$). The left-moving part, on the other hand, is obviously the bosonic CFT with $c = 24$ and spacetime gauge group $E_8^3$. There are in total 24 points of this type, differing in the choice of left-moving CFT based on the Niemeier lattices. Moreover, since the right-moving part is numerically equal to 24, there is an equal number of massive bosonic and fermionic level-matched states. Models of this type were considered in \cite{Kounnas:2008ft}, where they were named as MSDS models.

We can understand this in a different way by recalling that the Scherk-Schwarz reduction is constructed by using a half-shift along a compactification circle, but using T-duality we can place such a shift inside the gauge lattice; indeed in this way we can construct the different ten-dimensional non-supersymmetric heterotic strings with rank 16. Consider for example the $E_8 \times SO(16)$ string obtained from the $E_8 \times E_8$ supersymmetric heterotic string through an orbifold with $g = T(-1)^F$ where $T = e^{2\pi i \delta}$ is the usual shift vector with $\delta = (1,0^7)$ acting on the second $E_8$ by selecting out its underlying $D_8$ sublattice. If we started instead from the heterotic string on $T^8$ with Narain lattice polarized to 
\begin{equation}
	\Gamma_N \simeq E_8 \oplus E_8 \oplus E_8 \oplus E_8(-1)\,,
\end{equation}
we could place the shift on either one of the positive definite $E_8$ factors, giving rise to expression \eqref{e8d8}, or instead put the shift on the negative definite $E_8$, producing the isomorphic lattice commented below \eqref{e8d8}. Both choices are related by T-duality as indeed the choice of a shift vector of this type is unique up to automorphism \cite{Ginsparg:1986wr}, and in this way we verify the isomorphism $D_8 \oplus E_8(-1)\simeq E_8 \oplus D_8(-1)$.

Putting the shift on the negative definite $E_8$ amounts to transforming the supersymmetric theory into the product of the meromorhic CFT based in the Niemeier lattice $3\, E_8$ and the fermionic CFT with current $8\, \mathfrak{su}_{2,2}$. This is equivalent to the procedure of compactifying and polarizing, but gives us a more general perspective. All the supersymmetric heterotic strings with which we are concerned here are obtained as the tensor product of a meromorphic bosonic CFT with $c = 24$ and the $E_8$ sCFT. As such, we always have the option of orbifolding asymmetrically with $g = T(-1)^F$ with $T$ the shift located in the right-moving CFT. 

In the case of the Scherk-Schwarz reduction of the supersymmetric heterotic strings, which corresponds to genus $\mathcal{A}_I$, these results imply that there are $273+24$ holomorphic factorization points in the moduli space for the 2D theory, with the last 24 corresponding to MSDS models. Similarly, for genus $\mathcal{B}_I$ we will have $174+17$ factorization points, with the latter 17 being also MSDS models. Thus we see in this way that a family of MSDS models is intimately connected with the usual Scherk-Schwarz reduction applied to heterotic strings with 16 supercharges, as we include the remaining genera $\mathcal{C},...,\mathcal{K}$. There is also the possibility of taking the left-moving CFT in two dimensions to be the Monster CFT but in that case there are no classical moduli and so there is no decompactified theory to which we could apply the usual Scherk-Schwarz reduction; structually, however, it is of the same type as those theories which do admit such an interpretation.  
 
The notion of MSDS model is more general than what we are considering here. We could also take the right-moving CFT to have a different current algebra, with the only requirement that its dimension is 24 (e.g. $3\, \mathfrak{su}_{3,3}$), or take a product of two such CFTs, giving rise to a type II MSDS model. The interested reader is encouraged to look at the original reference \cite{Kounnas:2008ft}. Orbifolds of these MSDS models were subsequently constructed in \cite{Florakis:2009sm}, where their marginal deformations and interpretation in terms of higher dimensional string compactifications were studied in a similar spirit to the present work. It would be very interesting to see to what extent these other theories can be decompactified without losing their structure and get an idea of their salient properties.

\subsection{Genera of type $IIa$, $IIb$ and $III$}
As for the remaining genera let us start by considering those of type $IIb$. From \cite{Hohn:2023auw} (see e.g. their Figure 4), the corresponding CFTs can be constructed from the bosonic ones by combining a shift with an outer automorphism. In the string construction this translates to an orbifold of the type $R T (-1)^F$, as we saw in section \ref{ss:nonsusychl} for genus $\mathcal{B}_{IIb}$. The list of genera reads (cf. Table 9 of \cite{Hohn:2023auw})
\begin{equation}
	\text{Type IIb genera:} ~~~~~ \mathcal{B}\,, ~~~ \mathcal{D}\,, ~~~ \mathcal{E}\,, ~~~ \mathcal{G}\,, ~~~ \mathcal{I}\,, ~~~ \mathcal{J}\,, ~~~ \mathcal{K}\,.
\end{equation}
Here $\mathcal{D}$, $\mathcal{J}$ and $\mathcal{K}$ have order doubling and no six dimensional construction is known, but they do exist in five dimensions as can be read from the results of \cite{Persson:2015jka}. In any case, the orders of these orbifolds are always even; they are 2, $2\times 2$, 4, 6, 8, $2\times 12$ and $2 \times 10$. Thus inclusion of the factor $(-1)^F$ does not alter the order of the orbifold $RT$ which would be used to obtain the supersymmetric theory. For the genera $\mathcal{E}_{IIb}$, $\mathcal{G}_{IIb}$ and $\mathcal{I}_{IIb}$, breaking supersymmetry is just as for genus $\mathcal{B}_{IIb}$ and we can be sure that the corresponding six dimensional theories exist (seven dimensional in the case of genus $\mathcal{G}_{IIb}$, corresponding to a non-supersymmetric $\mathbb{Z}_4$-heterotic triple). 

Moving on to genera of type $IIa$, we see from Table 8 of \cite{Hohn:2023auw} that the list is identical to the one above for genera of type $IIb$. In this case however the same arguments above do not apply. It is true that genera of type $IIb$ can be constructed by starting from the supersymmetric theories associated to the symbol for the genus. For example, we have constructed genus $\mathcal{B}_{IIb}$ as a Chaudhuri-Polchinski orbifold of the $O(16)\times O(16)$ string, but this is equivalent to orbifolding the CHL string using $T (-1)^F$ where $T$ is a shift vector sitting on the $E_8$ lattice underlying the gauge algebra $\mathfrak{e}_{8,2}$, roughly speaking $\delta = (1,0^7)$ (the resulting gauge algebra is $\mathfrak{so}_{16,2}$). This is the situation that is also valid for the genera $\mathcal{B}_{IIa}$, where the shift is instead $\delta = (0^6,\tfrac12, \tfrac12)$ (the resulting gauge algebra is $\mathfrak{e}_{7,2} \oplus \mathfrak{su}_{2,2}$). Thus in order for the associated string theories to exist we should be able to find alternative shift vectors in the underlying supersymmetric strings just as we can do for the CHL string. We leave this problem for future work. 

Finally, genera of type $III$ can be constructed just as those of type $IIb$ but omitting the shift $T$, as we know to be the case for the $E_8$ string. From Table 10 of \cite{Hohn:2023auw}, this procedure can be carried out for the genera
\begin{equation}
	\text{Type III genera (1):} ~~~~~ \mathcal{B}\,, ~~~ \mathcal{D}\,, ~~~ \mathcal{G}\,, ~~~  \mathcal{J}\,,
\end{equation}
and following the discussion above we can be sure that in six dimensions we will find the associated string theories to genera $\mathcal{B}_{III}$ and $\mathcal{G}_{III}$ (in fact respectively in ten and seven dimensions). There are however certain extra genera that Hohn and Möller have found which correspond to outer automorphisms applied on bosonic CFTs with reduced rank. These are labeled $\mathcal{M}_{III}$ to $\mathcal{T}_{III}$, and we will make no attempt here at connecting them to associated string theories in higher dimensions. We should note however that genera $\mathcal{S}_{III}$ and $\mathcal{T}_{III}$ correspond to two dimensional theories without moduli fields (they are related to the monster CFT) and so the question does not apply.

\section{Conclusions}\label{s:conc}
In this work we have made explicit the correspondence between various non-supersymmetric heterotic strings and families of fermionic CFTs with $c = 24$ recently classified in \cite{Hohn:2023auw}. We have focused our attention on theories with rank reduced by 8, which can be thought of as non-supersymmetric versions of the CHL string of \cite{Chaudhuri:1995bf}, but have given some arguments for why and how this correspondence should extend to many other cases. 

In line with the view that there are as many inequivalent strings as there are separate CFT genera, we have found that in two different cases, two natural supersymmetry breaking constructions give rise to theories which are in fact T-dual. We have seen in particular that the Chaudhuri-Polchinski orbifold of the $O(16)\times O(16)$ string is T-dual to a straightforward non-supersymmetric version of the CHL string, the latter of which exhibits a partition function which is at face value simpler to work with. This of course also means that the related theories lie at infinite distance limits on a shared classical moduli space. 

We hope that this classification effort will bring some clarity and focus to the study of non-supersymmetric heterotic strings (in particular their realizations as asymmetric orbifolds). We have seen that the four theories with rank reduced by 8 can be distinguished by how extremal tachyons appear in their classical moduli spaces, as well as the maximal dimension where they can be constructed. It is likely that there are other salient features which distinguish such theories, which will be relevant at the formal level given that two of them are generically tachyonic. However, these features might be preserved under further compactification or perhaps dimension-changing tachyon condensation processes \cite{Hellerman:2007zz,Kaidi:2020jla}. To this end, we plan to provide an in depth exploration of these classical moduli spaces and determine what kinds of gauge symmetries and matter content can arise. One could also ask how these theories behave at infinite distance (see \cite{Collazuol:2024kzl} for a recent study of this question in the supersymmetric CHL string, and \cite{Font:2021uyw,Cvetic:2021sjm} for finite distance enhancements).

\section*{Acknowledgements}

We thank Anamaria Font, Bernardo Fraiman and Salvatore Raucci for many useful and stimulating discussions. We also thank Mariana Graña and Salvatore Raucci for useful comments on the manuscript, and Bernardo Fraiman for collaboration on related subjects. An important part of this work was carried out at the Institut de Physique Théorique, CEA Saclay. This work was partly supported 
by the ERC Consolidator Grant 772408-Stringlandscape, a grant from the Simons Foundation (602883,CV), the DellaPietra Foundation and by the NSF grant PHY2013858.
 
\bibliographystyle{JHEP}
\bibliography{nonsusyhet}

\providecommand{\href}[2]{#2}\begingroup\raggedright\begin{thebibliography}{10}

\bibitem{Dixon:1986iz}
L.~J. Dixon and J.~A. Harvey, {\it {String Theories in Ten-Dimensions Without
  Space-Time Supersymmetry}},  {\em Nucl. Phys. B} {\bf 274} (1986) 93--105.

\bibitem{Alvarez-Gaume:1986ghj}
L.~Alvarez-Gaume, P.~H. Ginsparg, G.~W. Moore, and C.~Vafa, {\it {An O(16) x
  O(16) Heterotic String}},  {\em Phys. Lett. B} {\bf 171} (1986) 155--162.

\bibitem{Kawai:1986vd}
H.~Kawai, D.~C. Lewellen, and S.~H.~H. Tye, {\it {Classification of Closed
  Fermionic String Models}},  {\em Phys. Rev. D} {\bf 34} (1986) 3794.

\bibitem{Seiberg:1986by}
N.~Seiberg and E.~Witten, {\it {Spin Structures in String Theory}},  {\em Nucl.
  Phys. B} {\bf 276} (1986) 272.

\bibitem{Ginsparg:1986bx}
P.~H. Ginsparg, {\it {Comment on Toroidal Compactification of Heterotic
  Superstrings}},  {\em Phys. Rev. D} {\bf 35} (1987) 648.

\bibitem{Narain:1985jj}
K.~S. Narain, {\it {New Heterotic String Theories in Uncompactified Dimensions
  \ensuremath{<} 10}},  {\em Phys. Lett. B} {\bf 169} (1986) 41--46.

\bibitem{Narain:1986am}
K.~S. Narain, M.~H. Sarmadi, and E.~Witten, {\it {A Note on Toroidal
  Compactification of Heterotic String Theory}},  {\em Nucl. Phys. B} {\bf 279}
  (1987) 369--379.

\bibitem{Scherk:1979zr}
J.~Scherk and J.~H. Schwarz, {\it {How to Get Masses from Extra Dimensions}},
  {\em Nucl. Phys. B} {\bf 153} (1979) 61--88.

\bibitem{Ferrara:1987es}
S.~Ferrara, C.~Kounnas, and M.~Porrati, {\it {Superstring Solutions With
  Spontaneously Broken Four-dimensional Supersymmetry}},  {\em Nucl. Phys. B}
  {\bf 304} (1988) 500--512.

\bibitem{Kounnas:1989dk}
C.~Kounnas and B.~Rostand, {\it {Coordinate Dependent Compactifications and
  Discrete Symmetries}},  {\em Nucl. Phys. B} {\bf 341} (1990) 641--665.

\bibitem{Vinberg}
E.~B. Vinberg, {\it The groups of units of certain quadratic forms},  {\em Mat.
  Sb. (N.S.)} {\bf 87(129)} (1972) 18--36.

\bibitem{Ginsparg:1986wr}
P.~H. Ginsparg and C.~Vafa, {\it {Toroidal Compactification of
  Nonsupersymmetric Heterotic Strings}},  {\em Nucl. Phys. B} {\bf 289} (1987)
  414.

\bibitem{Itoyama:1986ei}
H.~Itoyama and T.~R. Taylor, {\it {Supersymmetry Restoration in the
  Compactified O(16) x O(16)-prime Heterotic String Theory}},  {\em Phys. Lett.
  B} {\bf 186} (1987) 129--133.

\bibitem{Goddard:1986bp}
P.~Goddard and D.~I. Olive, {\it {Kac-Moody and Virasoro Algebras in Relation
  to Quantum Physics}},  {\em Int. J. Mod. Phys. A} {\bf 1} (1986) 303.

\bibitem{Cachazo:2000ey}
F.~A. Cachazo and C.~Vafa, {\it {Type I' and real algebraic geometry}},
  \href{http://arxiv.org/abs/hep-th/0001029}{{\tt hep-th/0001029}}.

\bibitem{Fraiman:2023cpa}
B.~Fraiman, M.~Gra\~na, H.~Parra De~Freitas, and S.~Sethi, {\it
  {Non-Supersymmetric Heterotic Strings on a Circle}},
  \href{http://arxiv.org/abs/2307.13745}{{\tt arXiv:2307.13745}}.

\bibitem{Baykara:2022cwj}
Z.~K. Baykara, D.~Robbins, and S.~Sethi, {\it {Non-Supersymmetric AdS from
  String Theory}},  \href{http://arxiv.org/abs/2212.02557}{{\tt
  arXiv:2212.02557}}.

\bibitem{Chaudhuri:1995fk}
S.~Chaudhuri, G.~Hockney, and J.~D. Lykken, {\it {Maximally supersymmetric
  string theories in D \ensuremath{<} 10}},  {\em Phys. Rev. Lett.} {\bf 75}
  (1995) 2264--2267, [\href{http://arxiv.org/abs/hep-th/9505054}{{\tt
  hep-th/9505054}}].

\bibitem{Chaudhuri:1995bf}
S.~Chaudhuri and J.~Polchinski, {\it {Moduli space of CHL strings}},  {\em
  Phys. Rev. D} {\bf 52} (1995) 7168--7173,
  [\href{http://arxiv.org/abs/hep-th/9506048}{{\tt hep-th/9506048}}].

\bibitem{Narain:1986qm}
K.~S. Narain, M.~H. Sarmadi, and C.~Vafa, {\it {Asymmetric Orbifolds}},  {\em
  Nucl. Phys. B} {\bf 288} (1987) 551.

\bibitem{Bianchi:1991eu}
M.~Bianchi, G.~Pradisi, and A.~Sagnotti, {\it {Toroidal compactification and
  symmetry breaking in open string theories}},  {\em Nucl. Phys. B} {\bf 376}
  (1992) 365--386.

\bibitem{Font:2021uyw}
A.~Font, B.~Fraiman, M.~Gra\~na, C.~A. N\'u\~nez, and H.~Parra De~Freitas, {\it
  {Exploring the landscape of CHL strings on $T^d$}},
  \href{http://arxiv.org/abs/2104.07131}{{\tt arXiv:2104.07131}}.

\bibitem{Mikhailov:1998si}
A.~Mikhailov, {\it {Momentum lattice for CHL string}},  {\em Nucl. Phys. B}
  {\bf 534} (1998) 612--652, [\href{http://arxiv.org/abs/hep-th/9806030}{{\tt
  hep-th/9806030}}].

\bibitem{Nakajima:2023zsh}
S.~Nakajima, {\it {New non-supersymmetric heterotic string theory with reduced
  rank and exponential suppression of the cosmological constant}},
  \href{http://arxiv.org/abs/2303.04489}{{\tt arXiv:2303.04489}}.

\bibitem{Schellekens:1992db}
A.~N. Schellekens, {\it {Meromorphic C = 24 conformal field theories}},  {\em
  Commun. Math. Phys.} {\bf 153} (1993) 159--186,
  [\href{http://arxiv.org/abs/hep-th/9205072}{{\tt hep-th/9205072}}].

\bibitem{Fraiman:2022aik}
B.~Fraiman and H.~Parra De~Freitas, {\it {Unifying the 6D $ \mathcal{N} $ = (1,
  1) string landscape}},  {\em JHEP} {\bf 02} (2023) 204,
  [\href{http://arxiv.org/abs/2209.06214}{{\tt arXiv:2209.06214}}].

\bibitem{Harrison:2021gnp}
S.~M. Harrison, N.~M. Paquette, D.~Persson, and R.~Volpato, {\it {BPS Algebras
  in 2D String Theory}},  {\em Annales Henri Poincare} {\bf 23} (2022), no.~10
  3667--3752, [\href{http://arxiv.org/abs/2107.03507}{{\tt arXiv:2107.03507}}].

\bibitem{Persson:2015jka}
D.~Persson and R.~Volpato, {\it {Fricke S-duality in CHL models}},  {\em JHEP}
  {\bf 12} (2015) 156, [\href{http://arxiv.org/abs/1504.07260}{{\tt
  arXiv:1504.07260}}].

\bibitem{Hohn:2023auw}
G.~H\"ohn and S.~M\"oller, {\it {Classification of Self-Dual Vertex Operator
  Superalgebras of Central Charge at Most 24}},
  \href{http://arxiv.org/abs/2303.17190}{{\tt arXiv:2303.17190}}.

\bibitem{BoyleSmith:2023xkd}
P.~Boyle~Smith, Y.-H. Lin, Y.~Tachikawa, and Y.~Zheng, {\it {Classification of
  chiral fermionic CFTs of central charge $\le 16$}},
  \href{http://arxiv.org/abs/2303.16917}{{\tt arXiv:2303.16917}}.

\bibitem{Rayhaun:2023pgc}
B.~C. Rayhaun, {\it {Bosonic Rational Conformal Field Theories in Small Genera,
  Chiral Fermionization, and Symmetry/Subalgebra Duality}},
  \href{http://arxiv.org/abs/2303.16921}{{\tt arXiv:2303.16921}}.

\bibitem{Kounnas:2008ft}
C.~Kounnas, {\it {Massive Boson-Fermion Degeneracy and the Early Structure of
  the Universe}},  {\em Fortsch. Phys.} {\bf 56} (2008) 1143--1156,
  [\href{http://arxiv.org/abs/0808.1340}{{\tt arXiv:0808.1340}}].

\bibitem{Witten:1997bs}
E.~Witten, {\it {Toroidal compactification without vector structure}},  {\em
  JHEP} {\bf 02} (1998) 006, [\href{http://arxiv.org/abs/hep-th/9712028}{{\tt
  hep-th/9712028}}].

\bibitem{deBoer:2001wca}
J.~de~Boer, R.~Dijkgraaf, K.~Hori, A.~Keurentjes, J.~Morgan, D.~R. Morrison,
  and S.~Sethi, {\it {Triples, fluxes, and strings}},  {\em Adv. Theor. Math.
  Phys.} {\bf 4} (2002) 995--1186,
  [\href{http://arxiv.org/abs/hep-th/0103170}{{\tt hep-th/0103170}}].

\bibitem{Acharya:2022shu}
B.~S. Acharya, G.~Aldazabal, A.~Font, K.~Narain, and I.~G. Zadeh, {\it
  {Heterotic strings on $ \mathbbm{T} ^{3}$/\ensuremath{\mathbb{Z}}$_{2}$,
  Nikulin involutions and M-theory}},  {\em JHEP} {\bf 09} (2022) 209,
  [\href{http://arxiv.org/abs/2205.09764}{{\tt arXiv:2205.09764}}].

\bibitem{Coudarchet:2021qwc}
T.~Coudarchet, E.~Dudas, and H.~Partouche, {\it {Geometry of orientifold vacua
  and supersymmetry breaking}},  {\em JHEP} {\bf 07} (2021) 104,
  [\href{http://arxiv.org/abs/2105.06913}{{\tt arXiv:2105.06913}}].

\bibitem{Mourad:2017rrl}
J.~Mourad and A.~Sagnotti, {\it {An Update on Brane Supersymmetry Breaking}},
  \href{http://arxiv.org/abs/1711.11494}{{\tt arXiv:1711.11494}}.

\bibitem{Basile:2023knk}
I.~Basile, A.~Debray, M.~Delgado, and M.~Montero, {\it {Global anomalies \&
  bordism of non-supersymmetric strings}},  {\em JHEP} {\bf 02} (2024) 092,
  [\href{http://arxiv.org/abs/2310.06895}{{\tt arXiv:2310.06895}}].

\bibitem{Kachru:2016ttg}
S.~Kachru, N.~M. Paquette, and R.~Volpato, {\it {3D String Theory and Umbral
  Moonshine}},  {\em J. Phys. A} {\bf 50} (2017), no.~40 404003,
  [\href{http://arxiv.org/abs/1603.07330}{{\tt arXiv:1603.07330}}].

\bibitem{Hohn:2017dsm}
G.~H\"ohn, {\it {On the Genus of the Moonshine Module}},
  \href{http://arxiv.org/abs/1708.05990}{{\tt arXiv:1708.05990}}.

\bibitem{Hohn:2020xfe}
G.~H\"ohn and S.~M\"oller, {\it {Systematic Orbifold Constructions of
  Schellekens' Vertex Operator Algebras from Niemeier Lattices}},
  \href{http://arxiv.org/abs/2010.00849}{{\tt arXiv:2010.00849}}.

\bibitem{Forgacs:1988iw}
P.~Forgacs, Z.~Horvath, L.~Palla, and P.~Vecsernyes, {\it {Higher Level
  {Kac-Moody} Representations and Rank Reduction in String Models}},  {\em
  Nucl. Phys. B} {\bf 308} (1988) 477--508.

\bibitem{NikulinVV1980ISBF}
V.~V. Nikulin, {\it Integral symmetric bilinear forms and some of their
  applications},  {\em Mathematics of the USSR. Izvestiya} {\bf 14} (1980),
  no.~1 103--167.

\bibitem{Florakis:2009sm}
I.~Florakis and C.~Kounnas, {\it {Orbifold Symmetry Reductions of Massive
  Boson-Fermion Degeneracy}},  {\em Nucl. Phys. B} {\bf 820} (2009) 237--268,
  [\href{http://arxiv.org/abs/0901.3055}{{\tt arXiv:0901.3055}}].

\bibitem{Hellerman:2007zz}
S.~Hellerman and I.~Swanson, {\it {A Stable vacuum of the tachyonic E(8)
  string}},  \href{http://arxiv.org/abs/0710.1628}{{\tt arXiv:0710.1628}}.

\bibitem{Kaidi:2020jla}
J.~Kaidi, {\it {Stable Vacua for Tachyonic Strings}},  {\em Phys. Rev. D} {\bf
  103} (2021), no.~10 106026, [\href{http://arxiv.org/abs/2010.10521}{{\tt
  arXiv:2010.10521}}].

\bibitem{Collazuol:2024kzl}
V.~Collazuol and I.~V. Melnikov, {\it {A twist at infinite distance in the CHL
  string}},  \href{http://arxiv.org/abs/2402.01606}{{\tt arXiv:2402.01606}}.

\bibitem{Cvetic:2021sjm}
M.~{Cvetic}, M.~{Dierigl}, L.~{Lin}, and H.~Y. {Zhang}, {\it {On the Gauge
  Group Topology of 8d CHL Vacua}},  {\em arXiv e-prints} (July, 2021)
  arXiv:2107.04031, [\href{http://arxiv.org/abs/2107.04031}{{\tt
  arXiv:2107.04031}}].

\end{thebibliography}\endgroup

\end{document}